\documentclass[pdflatex]{sn-jnl} 

\usepackage{graphicx}
\usepackage{multirow}
\usepackage{amsmath,amssymb,amsfonts,mathrsfs}
\usepackage{rotating}
\usepackage{appendix}
\usepackage{textcomp}
\usepackage{manyfoot}
\usepackage{booktabs}
\usepackage{etoolbox}
\usepackage{algorithm}
\usepackage{xcolor}
\usepackage{algorithmicx}
\usepackage{algpseudocode}
\usepackage{listings}
\usepackage{url}
\usepackage{adjustbox}
\usepackage{subcaption}  

\usepackage{geometry}
\usepackage{hypcap}
\usepackage{hyperref}
\usepackage{wrapfig}
\usepackage{amsthm}

\hypersetup{
  colorlinks=true,
  linkcolor=blue,
  citecolor=blue,
  urlcolor=blue
}

\title[Two-Faced Social Agents]{Two-Faced Social Agents: Context Collapse in Role-Conditioned Large Language Models}

\author*[1]{\fnm{Vikram} \sur{K. Suresh}}\email{krishnvv@ucmail.uc.edu}
\affil[1]{\orgdiv{Department of Computer Science}, \orgname{University of Cincinnati}, \city{Cincinnati}, \state{OH}, \country{USA}}

\abstract{As large language models (LLMs) gain prominence as social agents capable of simulating human behavior, recent work has focused on their distributional fidelity, the ability to reproduce aggregate patterns of human responses. Yet, the impact of role conditioning is unclear when task performance depends on role-specific cognitive constraints. In this study, we evaluate the persona fidelity of frontier LLMs, GPT-5, Claude Sonnet 4.5 and Gemini 2.5 Flash when assigned distinct socioeconomic personas performing scholastic assessment test (SAT) mathematics items and affective preference tasks. Across 15 distinct role conditions and three testing scenarios, GPT-5 exhibited complete contextual collapse and adopted a singular identity towards optimal responses (PERMANOVA $p=1.000$, $R^2=0.0004$), while Gemini 2.5 Flash showed partial collapse ($p=0.120$, $R^2=0.0020$). Claude Sonnet 4.5 retained limited but measurable role-specific variation on the SAT items (PERMANOVA $p<0.001$, $R^2=0.0043$), though with inverted SES-performance relationships where low-SES personas outperformed high-SES personas ($\eta^2$ = 0.15-0.19 in extended replication). However, all models exhibited distinct role-conditioned affective preference (average $d$ = 0.52-0.58 vs near zero separation for math), indicating that socio-affective variation can reemerge when cognitive constraints are relaxed. These findings suggest that distributional fidelity failure originates in task-dependent contextual collapse: optimization-driven identity convergence under cognitive load combined with impaired role-contextual understanding. Realistic social simulations may require embedding contextual priors in the model's post-training alignment and not just distributional calibration to replicate human-like responses. Beyond simulation validity, these results have implications for survey data integrity, as LLMs can express plausible demographic variation on preference items while failing to maintain authentic reasoning constraints.}


\begin{document}
\maketitle

\section{Introduction}\label{sec1}
The use of Large Language Models (LLMs) as social agents has gained significant traction across psychology, economics and computational social science \cite{Horton2023}. Given the vast training corpora of these models, these models are often assumed to possess the capacity to approximate human behavior across diverse contexts \cite{Grossmann2023, Ziems2023}. Recent studies have extended this assumption to experimental designs, employing LLMs as engines for role-conditioned populations, i.e., virtual participants prompted to adopt social identities or personas intended to shape their responses and interactions \cite{Park2023, Park2024}. Further, Manning and Horton (2025) proposed the use of ``General Social Agents'' (GSAs), a theory-grounded framework for constructing agents using LLMs that generate human-consistent distributions of choices across novel multi-agent strategic games \cite{Manning2025}.

While these efforts demonstrate that LLMs can reproduce aggregate human-like behavior most evaluation benchmarks are primarily focused on population-level distributional fidelity \cite{Hu2025, Argyle2023}. Which means, success is defined by how closely model response distributions resemble human samples. Yet, the high distributional fidelity may not suggest the LLM has successfully internalized the contextual socioeconomic role constraints that drive heterogeneous human behavior \cite{Wylie2003}. Recent work by Wang et al. (2025) raised a fundamental concern about this issue, demonstrating that LLMs systematically misportray and flatten demographic identities when used as synthetic participants. Their findings suggest models fail to capture within-group heterogeneity in static survey responses and produce homogenized representations that obscure lived experiences \cite{Wang2025}. However, this analysis was limited to static surveys and low cognitive preference elicitation tasks.

Benchmarks such as SimBench operationalize the goal of evaluating LLMs' ability to simulate human behavior by comparing model-generated response distributions to empirical human data across moral, social and reasoning tasks \cite{Hu2025}. Yet several studies now indicate that even when LLMs achieve near human aggregate performance, they diverge mechanistically from human cognition under stress and incentive variations. For instance, in prisoner's dilemma games, LLMs converge towards an immediate reward equilibrium rather than long-term strategic gains often observed in human subjects \cite{Akata2025}. Similarly, Kim et al. demonstrate that persona prompts are a ``double-edged sword,'' while occasionally improving performance, they often degrade when cognitive demands increase \cite{Kim2024}. Their proposed \textit{Jekyll\&Hyde} combining a persona-prompt and a neutral prompt with an evaluator selecting the better answer illustrates that such approaches improve accuracy but effectively abandon continuous persona enactment \cite{Kim2024}.

Beyond performance degradation, Wang et al. (2025) and Gupta et al. (2023) further demonstrate that when tasked with simulating social identities, previous generation of LLMs (GPT-3.5 and 4 class) often produce flattened variance and sometimes biased misrepresentations of politically sensitive groups \cite{KinderWinter2001, Sheng2021, Gupta2023, Wan2023,Wang2025}. Taken together, these findings suggest distributional realism is insufficient to ensure role-contextual fidelity while maintaining distinct contextual selves while reasoning. Manning and Horton (2025) similarly argue that durable social agents must embed contextual priors from theory driven frameworks rather than solely relying on distributional calibration \cite{Manning2025}.

The stakes of these limitations extend beyond academic validity. Recent concerns have emerged about LLM infiltration of survey panels, where bad actors could manipulate public opinion measurement at minimal cost \cite{Westwood2025}. Understanding when and how LLMs can sustain persona fidelity versus when they collapse is therefore critical not only for simulation validity but for survey data integrity.

This study formalizes that hypothesis empirically. We test whether LLMs can maintain role-conditioning under objective cognitive load (SAT reasoning) versus subjective affective preference reporting. Socioeconomic personas are used in place of protected identities to avoid demographic misportrayal \cite{Wang2025}. By connecting micro-level collapse to macro-level distributional fidelity failure, we demonstrate that realistic social simulation requires not merely matching population frequencies but sustaining consistent contextual selves under reasoning. This extends the logic presented by Wang et al. (2025) by examining how models sustain or collapse socioeconomic role-conditioning during complex reasoning. Where prior work quantified representational harms, we examine their cognitive analogue, the erosion of contextual selfhood as optimization drives models towards deterministic uniformity. This selective collapse reveals what we term as 'two-faced social agents,' models that appear to embody diverse identities in low-stakes contexts but revert to optimization driven homogeneity when demands increase.

To evaluate this mechanism, we selected state-of-the-art LLMs, GPT-5, Claude Sonnet 4.5, and Gemini 2.5 Flash \cite{OpenAI2025GPT5, Anthropic2025Sonnet45, GoogleDeepMind2025Gemini2.5Flash}. These LLMs were selected based on their advanced capabilities and being the frontier LLMs as of late 2025. OpenAI's GPT-5 emphasizes reduced hallucinations and stronger factual reasoning \cite{OpenAI2025GPT5}. Given its markets share and influence, GPT-5 is the most likely vector for legitimate simulations and potential misuse in survey contexts \cite{TullyEtAl2025,Westwood2025}. Hence we include it as a benchmark for role-conditioning fidelity and hypothesize it may exhibit contextual collapse under cognitive load. Claude Sonnet 4.5 incorporates Constitutional AI principles emphasizing harmlessness and scenario awareness that can influence alignment behavior even without explicit prompting \cite{Anthropic2025Sonnet45, Bai2022, Askell2021}. Gemini 2.5 Flash on the other hand, was optimized for multi-modal reasoning and tool usage to align better with human preference data \cite{GoogleDeepMind2025Gemini2.5Flash}. Each of these models use unique training and alignment strategies, providing a diverse set of architectures for evaluating role-conditioning fidelity. The models are also significant advances from the previous architectures examined by Wang et al. (2025) and Gupta et al. (2023), with sophisticated alignment techniques that may mitigate prior limitations \cite{Gupta2023, Wang2025}. If persona fidelity fails to emerge in the current generation of LLMs, it would suggest a fundamental limitation in current alignment paradigms.

We evaluate each model on two tasks designed to contrast deterministic cognitive optimization with expressive contextual reasoning as show in Table. \ref{tab:experiment_design}.
\begin{itemize}
  \item \textbf{Scholastic Assessment Test (SAT) Mathematics Items:} A set of multiple choice and numerical reasoning problems from the SAT mathematics section. These items have a single correct outcome, forcing the model to optimize for accuracy under role-specific cognitive constraints.
  \item \textbf{Affective Preference Tasks:} A series of subjective preference questions from established economic and social psychology literature that permit a range of acceptable responses. This enables the model to express role-conditioned affective variation across their assigned socioeconomic persona.
\end{itemize}

Cognitive loading thus serves as a stringent test of role-conditioning fidelity, if a model maintains distinct reasoning patterns across personas despite fixed correction answers, it demonstrates genuine contextual grounding. Conversely, uniform responses across roles indicate contextual collapse driven by optimization pressures. The affective preference tasks provides a complementary baseline, revealing whether role-conditioned variation can reemerge when cognitive constraints are relaxed. Building on prior concerns by Wang et al., we study potential limitations in the current frontier LLMs' ability to sustain role-conditioned reasoning. First, contextual misalignment, whereby models trained for optimization reproduce what a generic solver would infer rather than reason as that role. Second, contextual flattening, where optimization pressures erases heterogeneity across role-conditioned personas, to produce a uniform reasoning style. Third, contextual essentialism, in which the role attributes are surface prompts rather than internalized identities during reasoning.

\begin{table*}[ht!]
  \centering
  \caption{\textbf{Experimental design overview:}
    Each large language model (LLM) was evaluated across two task types:
    (1) SAT mathematics reasoning and (2) affective preference reporting.
    The primary experiment included 15 socioeconomic personas and three testing scenarios.
    A replication with 45 personas was conducted for Claude Sonnet~4.5.
    Both deterministic ($T = 0.0$) and stochastic ($T = 0.6$) decoding were used in the main experiment,
    while only deterministic decoding was used for the replication.
    Preference tasks were assumed scenario-stable.}
  \label{tab:experiment_design}
  \vspace{0.6em}
  \begin{adjustbox}{max width=\textwidth}
    \footnotesize
    \begin{tabular}{@{}p{2.5cm}p{2.8cm}cp{2.8cm}p{4.2cm}@{}}
      \toprule
      \textbf{Task Type}                         & \textbf{Items} & \textbf{Agents} & \textbf{Models Tested} & \textbf{Notes / Constraints}                                            \\
      \midrule

      \textbf{SAT Math (main)}                   &
      28 valid questions                         &
      15                                         &
      GPT-5, Claude Sonnet~4.5, Gemini~2.5~Flash &
      Two decoding settings: ($T=0.0$) and ($T=0.6$). Each persona evaluated across all three scenarios (optimal, moderate stress, challenging).                                       \\[0.8em]

      \textbf{Preference Task}                   &
      16 + 1 validation                          &
      15                                         &
      GPT-5, Claude Sonnet~4.5, Gemini~2.5~Flash &
      Affective questions assessing subjective preferences. Validation item repeated agent self-name to ensure identity recall.                                                        \\[0.8em]

      \textbf{SAT Math (replication)}            &
      54 questions                               &
      45                                         &
      Claude Sonnet~4.5 only                     &
      Full replication using expanded persona set (45$\times$3). Used for high-resolution analysis of role fidelity variance across SES strata. Only deterministic decoding ($T=0.0$). \\

      \textbf{Preference Task (replication)}     &
      16 + 1 validation                          &
      45                                         &
      Claude Sonnet~4.5 only                     &
      Affective questions assessing subjective preferences. Validation item repeated agent self-name to ensure identity recall.                                                        \\
      \bottomrule
    \end{tabular}
  \end{adjustbox}
\end{table*}

\section{Results}\label{sec2}
In this section, we present the results from the main experiment evaluating GPT-5, Claude Sonnet 4.5 and Gemini 2.5 Flash across the SAT mathematics reasoning and affective preference tasks. A total of 28 questions from the SAT mathematics section were used after filtering for valid response formats. Each model was prompted to simulate 15 distinct socioeconomic personas across three testing scenarios (optimal, moderate stress, challenging) resulting in 1260 unique SAT responses per model. Additionally, each model responded to 16 affective preference questions plus one validation item across the same personas, yielding 240 preference responses per model.

\subsection{Main Experiment}
For the SAT mathematics items, we first evaluated overall accuracy across models and each scenario. As shown in Figure \ref{fig:sat_accuracy_ses_scenarios}, GPT-5 achieved the highest overall accuracy with 100\% correct responses across all personas and scenarios. Gemini 2.5 Flash followed with an average accuracy of 100\% in all three scenarios. Claude Sonnet 4.5 exhibited variable accuracies across personas but this did not extend across scenarios, with an accuracy of 95\% for low-SES, 95.54\% for middle-SES, and 91.07\% for high-SES personas all three scenarios. Further, we examined the SES-based differences for Claude Sonnet 4.5 using one-way ANOVA across the three SES groups (Low: N=5, Middle: N=4, High: N=6). The omnibus test revealed consistent patterns across all three scenarios (F(2,12)=2.21, p=0.15, $\eta^2=0.27$), with Low-SES personas showing higher accuracy (M=95.0\%, SD=3.2) than High-SES personas (M=91.1\%,SD=4.4; Cohen's $d$=1.03 for pairwise comparison). While this represented a large effect size by conventional standards, the comparison did not reach statistical significance, potentially due to limited statistical power with small group sizes. As a result, we conducted an extended replication with 45 personas to increase statistical power and determine whether the observed pattern reflected genuine SES-based trait embodiment or sampling variability.

\begin{center}
  \includegraphics[width=0.9\textwidth]{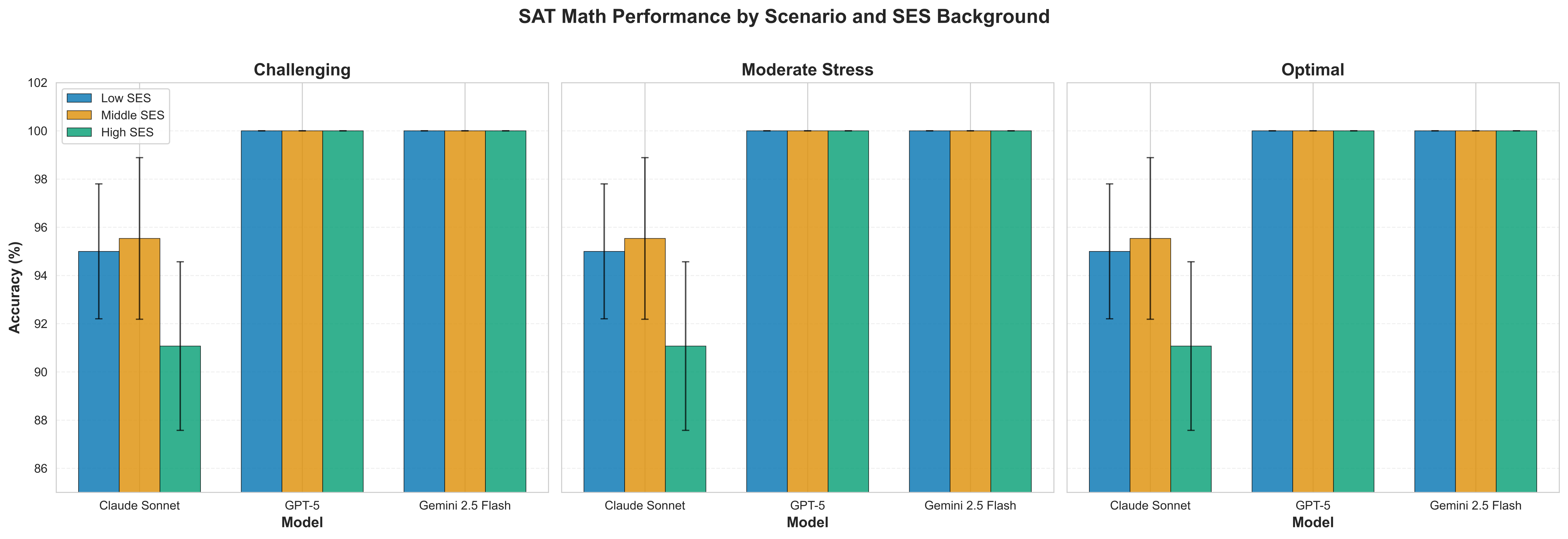}
  \captionof{figure}{SAT mathematics accuracy across socioeconomic personas and testing scenarios for each model.
    (a) GPT-5 exhibited complete contextual collapse with uniform accuracy across SES groups and scenarios.
    (b) Gemini 2.5 Flash also collapses under all scenarios and SES groups.
    (c) Claude Sonnet 4.5 retained measurable SES-based accuracy differences across all scenarios prompting extended validation.}
  \label{fig:sat_accuracy_ses_scenarios}
\end{center}

The accuracy patterns reveal the first dimension of contextual collapse. When faced with single answer, models eliminate SES-based heterogeneity entirely or show systematic inversion. This contrasts with the preference task results, suggesting that cognitive load fundamentally alters role-conditioning fidelity. If models truly internalized socioeconomic personas, we would observe SES-based accuracy variation across both domains. Conversely, if collapse is optimization driven, we would expect task-dependent trait expression to emerge only when correctness pressures are absent.

\begin{figure}[ht]
  \centering
  \begin{subfigure}[t]{0.48\textwidth}
    \centering
    \includegraphics[width=\textwidth]{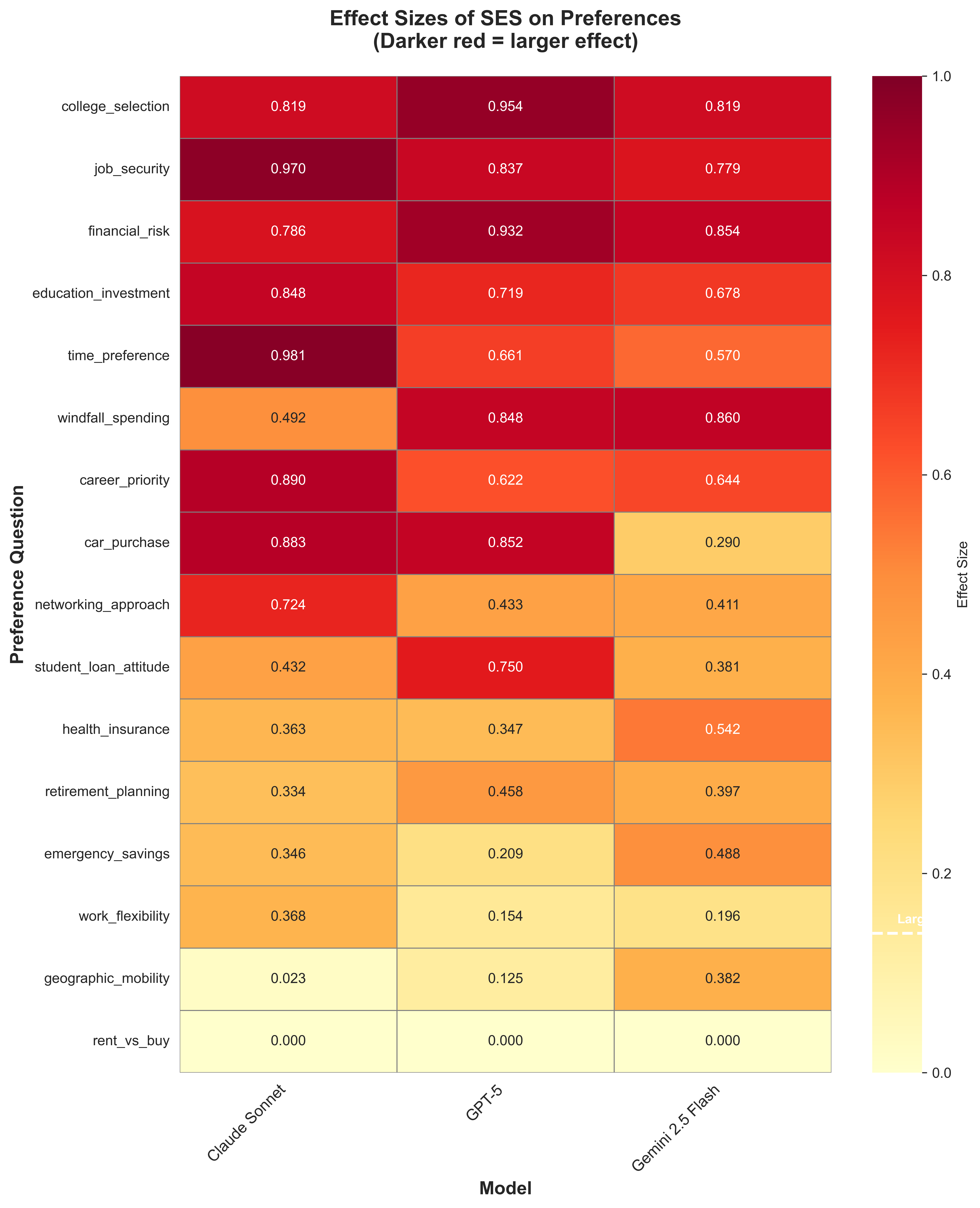}
    \caption{Effect size heatmap ($\epsilon^2$ / Cramer's $V$)}
    \label{fig:pref_effectsize_heatmap}
  \end{subfigure}
  \hfill
  \begin{subfigure}[t]{0.48\textwidth}
    \centering
    \includegraphics[width=\textwidth]{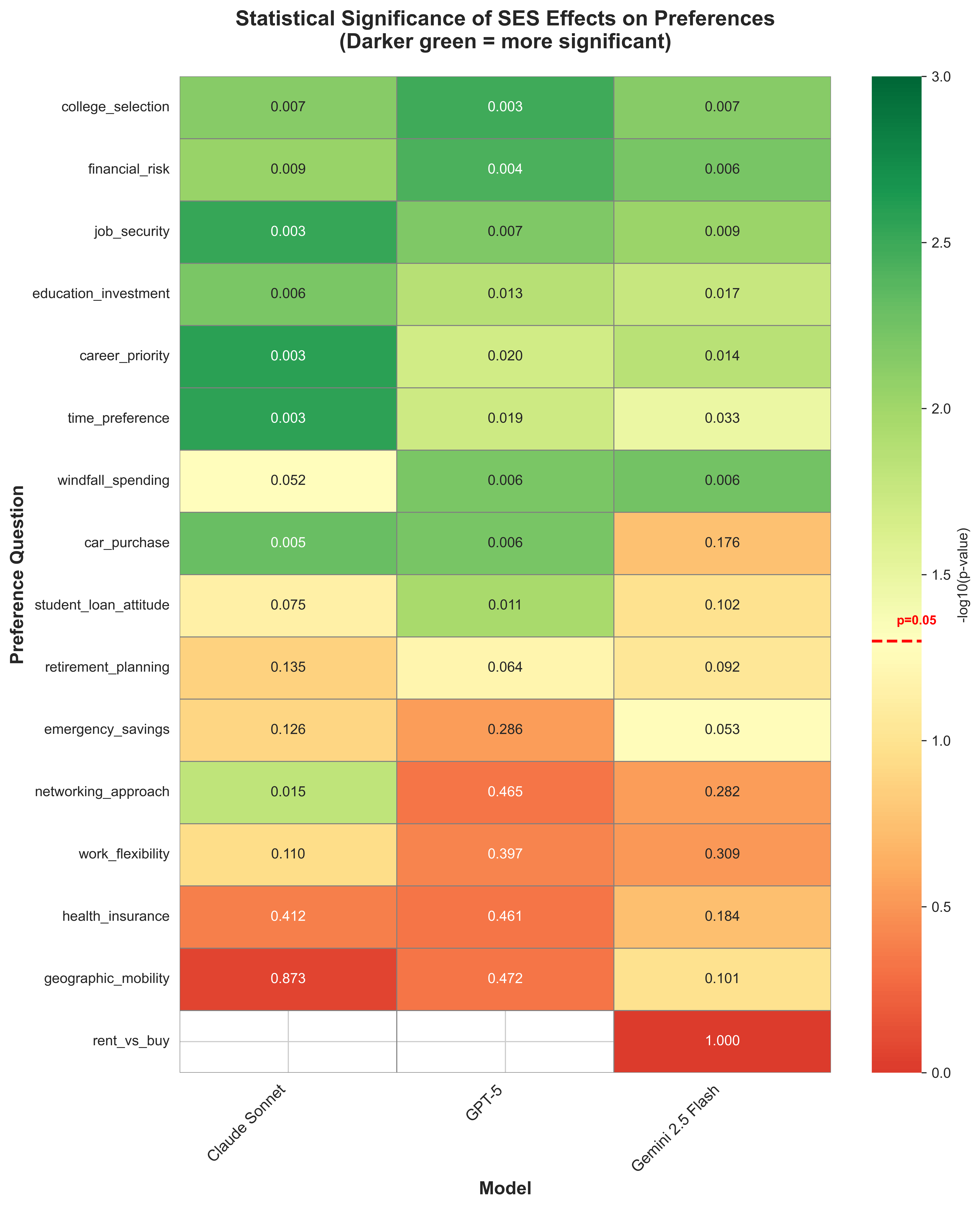}
    \caption{Statistical significance heatmap ($p$-values)}
    \label{fig:pref_pvalue_heatmap}
  \end{subfigure}

  \caption{Preference task SES analysis across 16 economic items and three models.
    (a) Effect sizes for ordinal and categorical preference dimensions.
    (b) Corresponding $p$-value heatmap showing the robustness and direction of SES associations.}
  \label{fig:preference_heatmaps}
\end{figure}

Therefore, we examined the affective preference task when the cognitive constraints were relaxed. Figure \ref{fig:preference_heatmaps} presents the effect size and statistical significance heatmaps across the 16 preference items for each model. These effect sizes were computed using Cramer's V for categorical items and $\epsilon^2$ for ordinal items (see Section \ref{sec4}). Models exhibited substantial SES-based preference variation across multiple items with the average effect size across all items for Claude Sonnet 4.5 was $d=0.58$, for GPT-5 it was $d=0.56$, and for Gemini 2.5 Flash it was $d=0.52$. Notably, models do not have statistically significant SES-based differences on all preference items. Items such as student loan attitude, retirement planning, emergency savings, networking approach, work flexibility, health insurance, geographic mobility and rent vs.\ buy preferences showed no significant SES differences across all models ($p>0.05$). However, other items such as risk tolerance, time preference, college choice, career priorities, windfall spending and car purchase decisions exhibited strong SES-based variation ($p<0.05$) across all models. Since all models passed the identity validation item (repeating their assigned persona name), we can conclude that the models successfully recalled their socioeconomic personas during the preference tasks.

The collapse under cognitive load while preserving under subjective preference reveals the two-faced nature of current social agents. Models demonstrate socioeconomic variation when expressing preferences about risk, time and consumption (average $d=0.55$), but this variation vanishes when optimizing for correctness. This pattern suggests models have not internalized the personas as stable identities but rather apply them selectively on task affordances.

\begin{figure}[ht]
  \centering

  \begin{subfigure}[t]{0.48\linewidth}
    \centering
    \includegraphics[width=\linewidth]{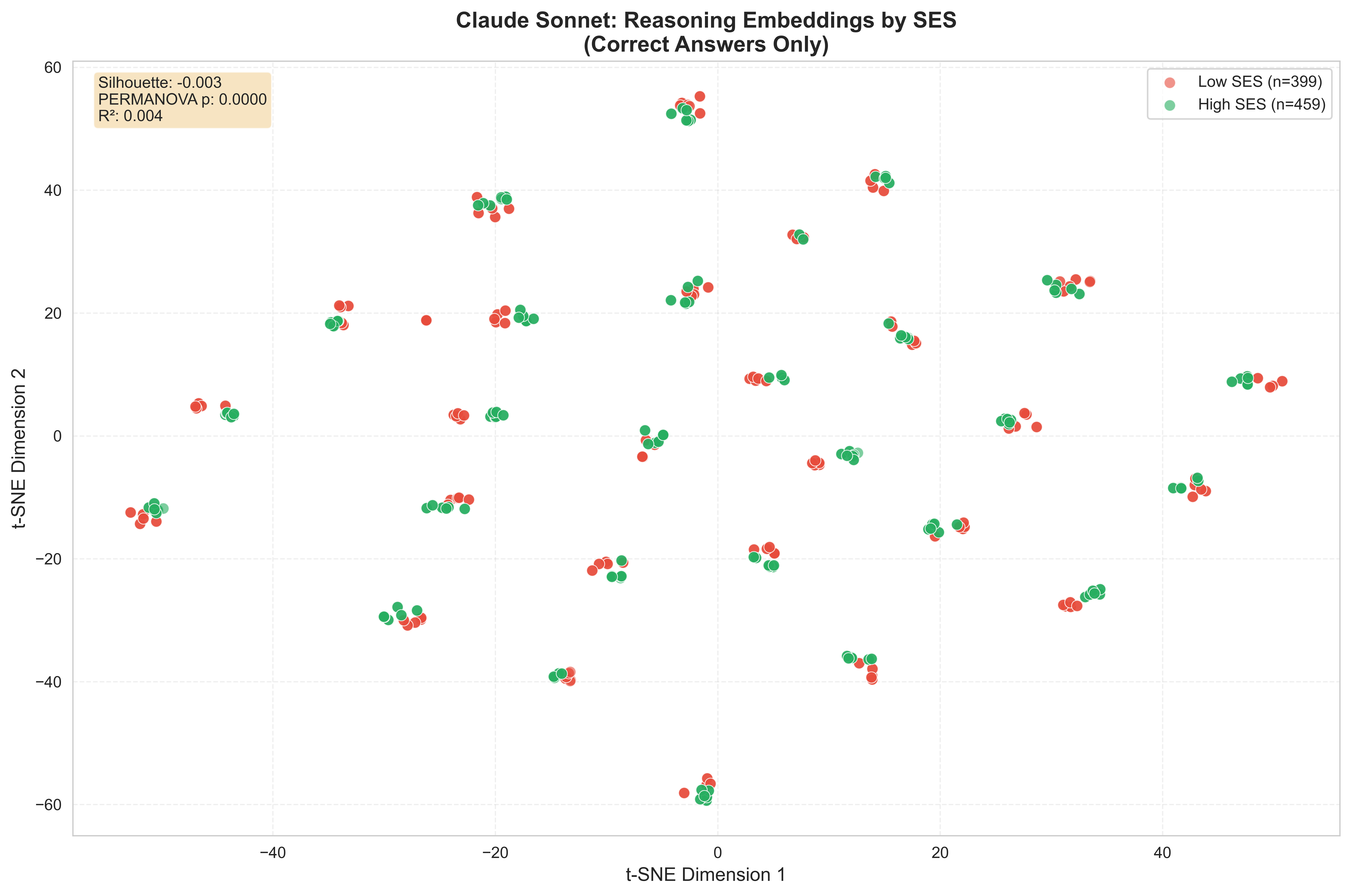}
    \caption*{\textbf{Claude Sonnet}\\\small Minor SES separation}
  \end{subfigure}
  \hspace{0.01\linewidth}
  \begin{subfigure}[t]{0.48\linewidth}
    \centering
    \includegraphics[width=\linewidth]{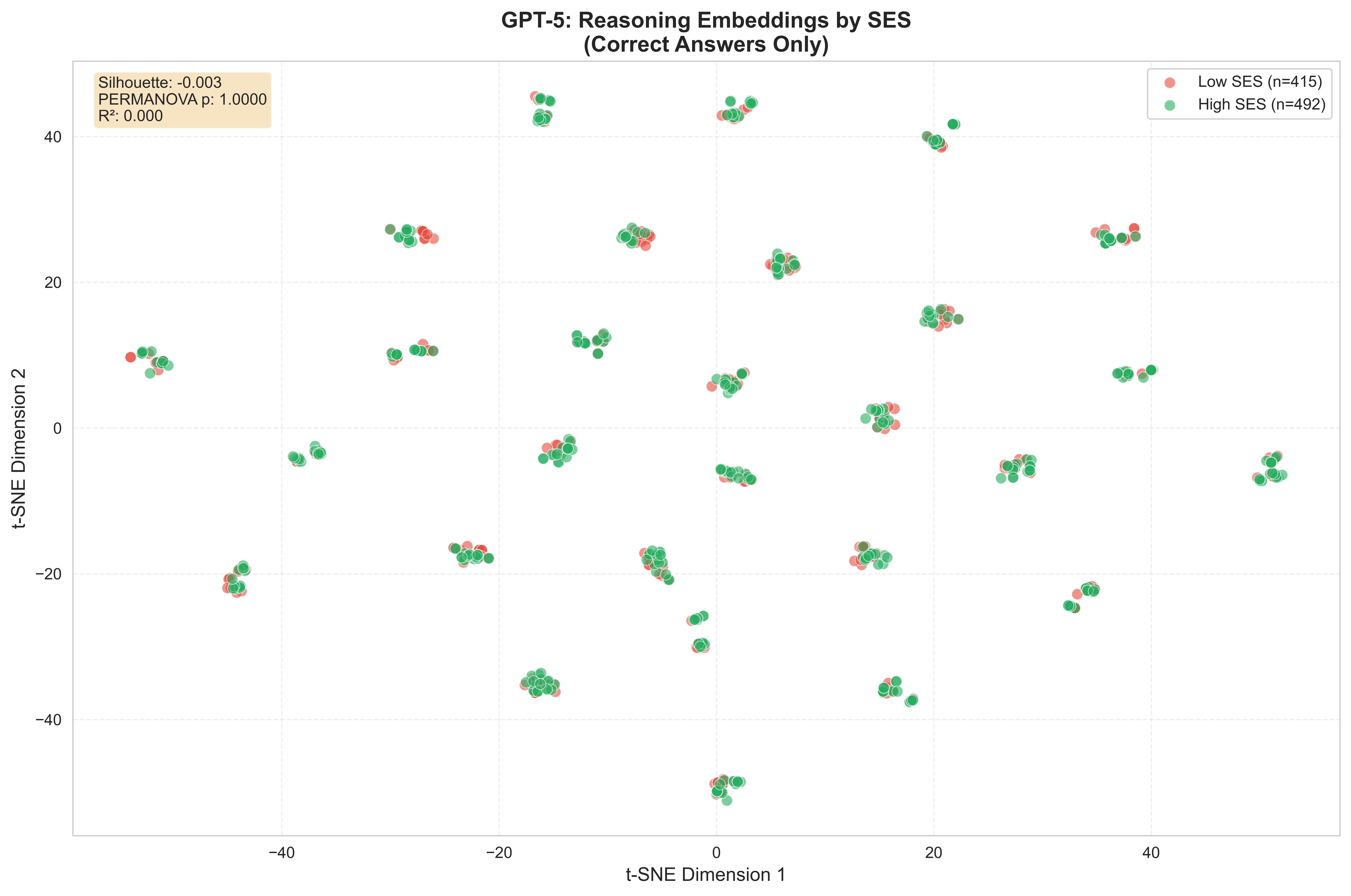}
    \caption*{\textbf{GPT-5}\\\small No SES separation}
  \end{subfigure}
  \hspace{0.01\linewidth}
  \begin{subfigure}[t]{0.48\linewidth}
    \centering
    \includegraphics[width=\linewidth]{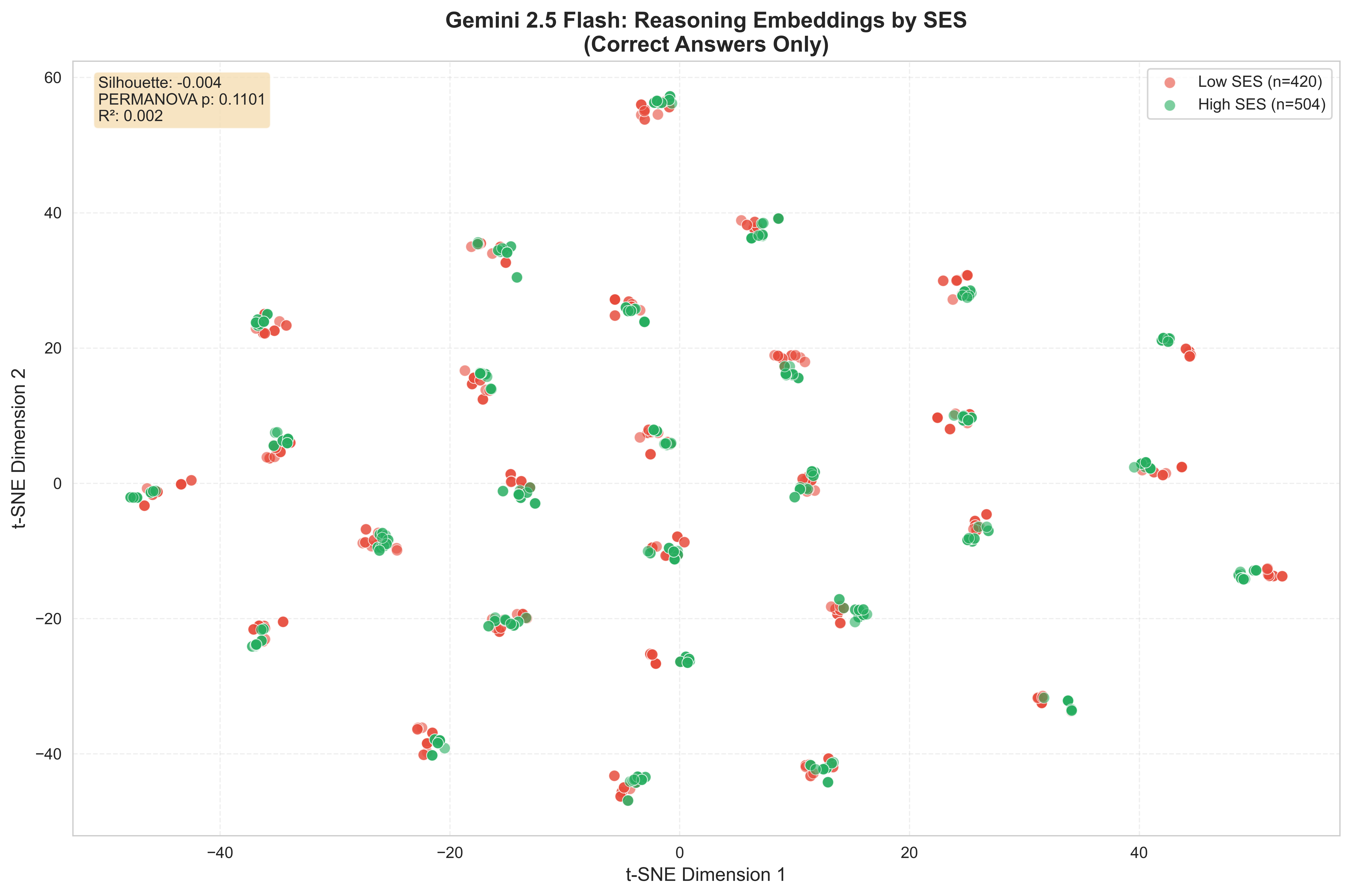}
    \caption*{\textbf{Gemini 2.5 Flash}\\\small Minor SES separation}
  \end{subfigure}

  \caption{
    t-SNE projections of reasoning embeddings from correct SAT solutions.
    Claude Sonnet and Gemini 2.5 exhibit slight SES-structured drift, while
    GPT-5 shows complete suppression of SES structure.
  }
  \label{fig:ses_clustering_three_models}
\end{figure}

Clustering analysis of the reasoning embeddings from correct SAT solutions further illustrated the degree of contextual collapse across models. Figure \ref{fig:ses_clustering_three_models} presents t-SNE projections of the reasoning embeddings colored by SES group for each model. GPT-5 exhibited complete collapse with no discernible SES-based clustering, a silhouette score of $-0.0034$ and PERMANOVA $p=1.000$, $R^2=0.0004$. Gemini 2.5 Flash showed minor SES-structured drift with a silhouette score of $-0.0038$ and PERMANOVA $p=0.120$, $R^2=0.0020$. Claude Sonnet 4.5 retained limited but measurable SES-based clustering with a silhouette score of $0.014$ and PERMANOVA $p<0.001$, $R^2=0.0043$. The latent representation for different personas are highly similar for GPT-5 and Gemini 2.5 Flash, indicating that the models converged towards a singular reasoning patterns across SES categories and can be seen in the overlapping t-SNE clusters.

\begin{figure}[ht]
  \centering

  \begin{subfigure}[t]{0.48\textwidth}
    \centering
    \includegraphics[width=\textwidth]{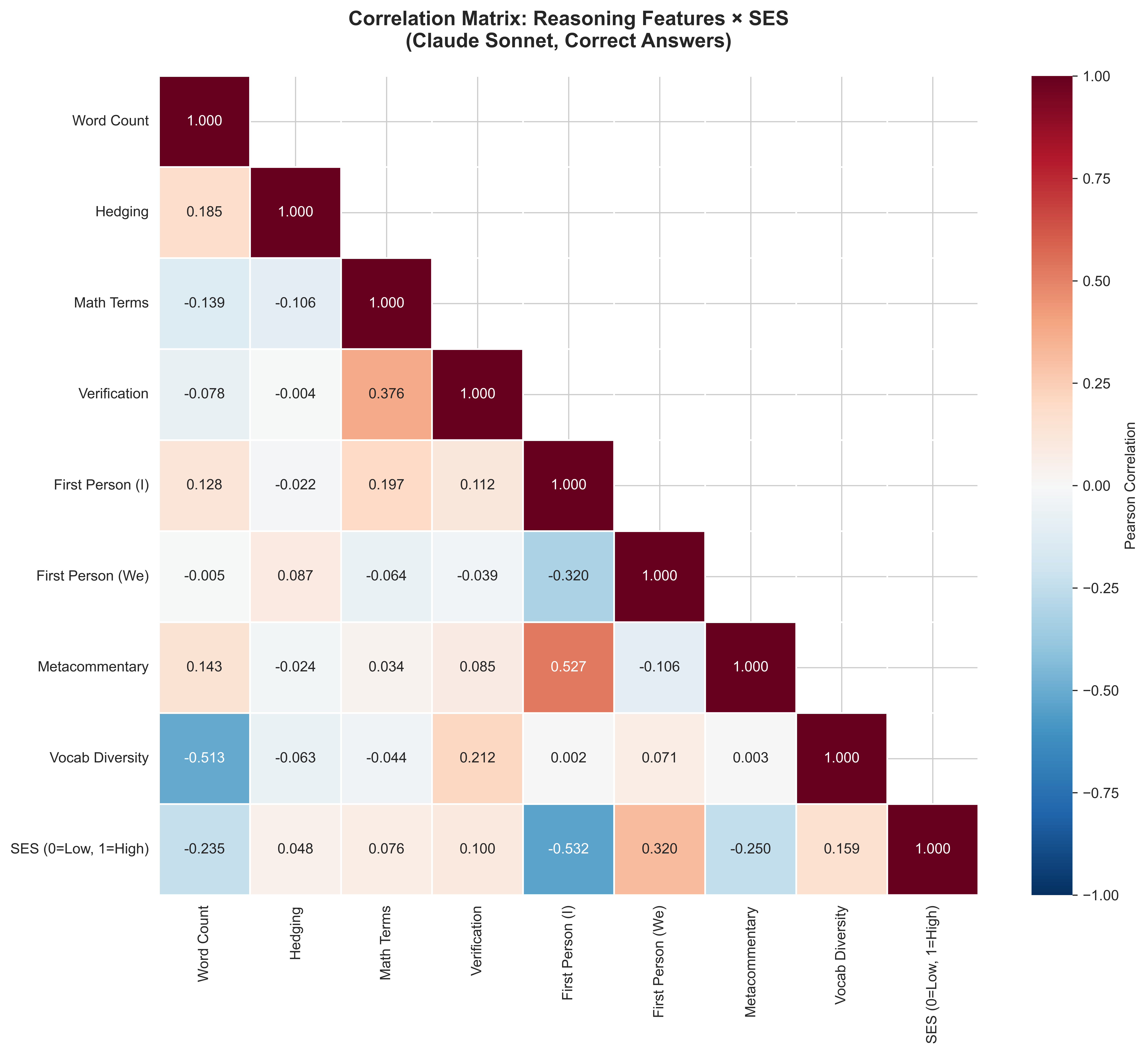}
    \caption{Linguistic feature correlation structure}
  \end{subfigure}
  \hfill
  \begin{subfigure}[t]{0.48\textwidth}
    \centering
    \includegraphics[width=\textwidth]{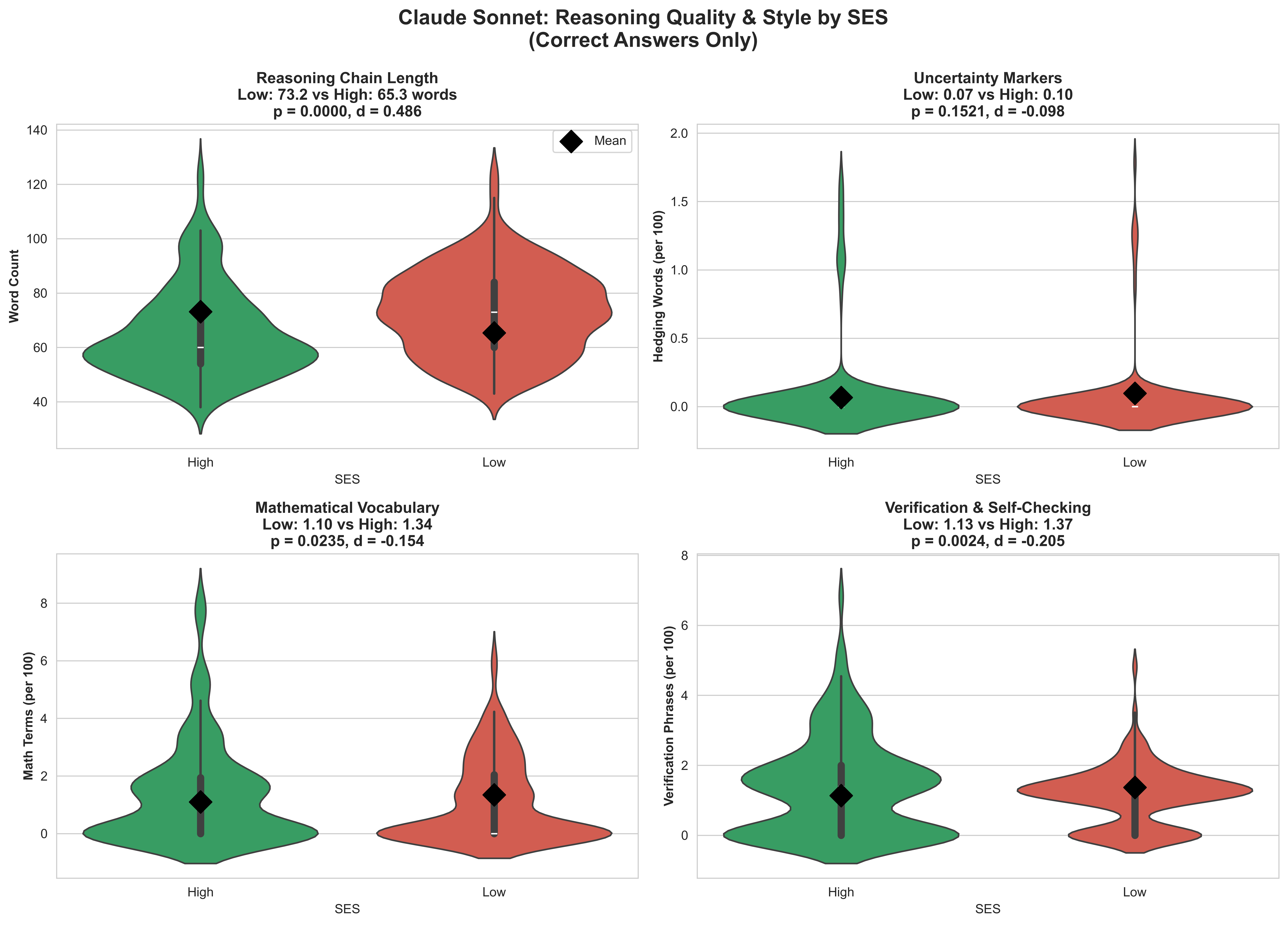}
    \caption{Top SES-differentiating linguistic features}
  \end{subfigure}

  \caption{Reasoning quality and linguistic structure in Claude compared to other models.}
  \label{fig:reasoning_linguistic_results}
\end{figure}

Further linguistic analysis of the reasoning outputs revealed distinct patterns in Claude Sonnet 4.5 compared to the other models. Figure \ref{fig:reasoning_linguistic_results}(a) presents the correlation matrix of linguistic features extracted from the reasoning outputs. First person pronoun usage (e.g., "I", "me") was negatively correlated with high-SES personas ($r=-0.532$), while other magnitudes $|r| \approx 0.1-0.2$ indicate weak correlations across features. This suggests that Claude Sonnet 4.5 is subtly modulating its linguistic style based on the assigned socioeconomic persona. Figure \ref{fig:reasoning_linguistic_results}(b) shows linguistic features that most strongly differentiate low-SES and high-SES personas. Features such as average sentence length ($d=0.49$), verification steps ($d=0.21$) and less mathematical vocabulary ($d=0.15$) were more prevalent in low-SES personas. Other markers including hedging, uncertainty and meta-commentary showed no significant SES differences, indicating variation likely reflects quantitative elaboration rather than qualitatively different reasoning strategies.

\begin{figure}[ht]
  \centering
  \includegraphics[width=0.85\textwidth]{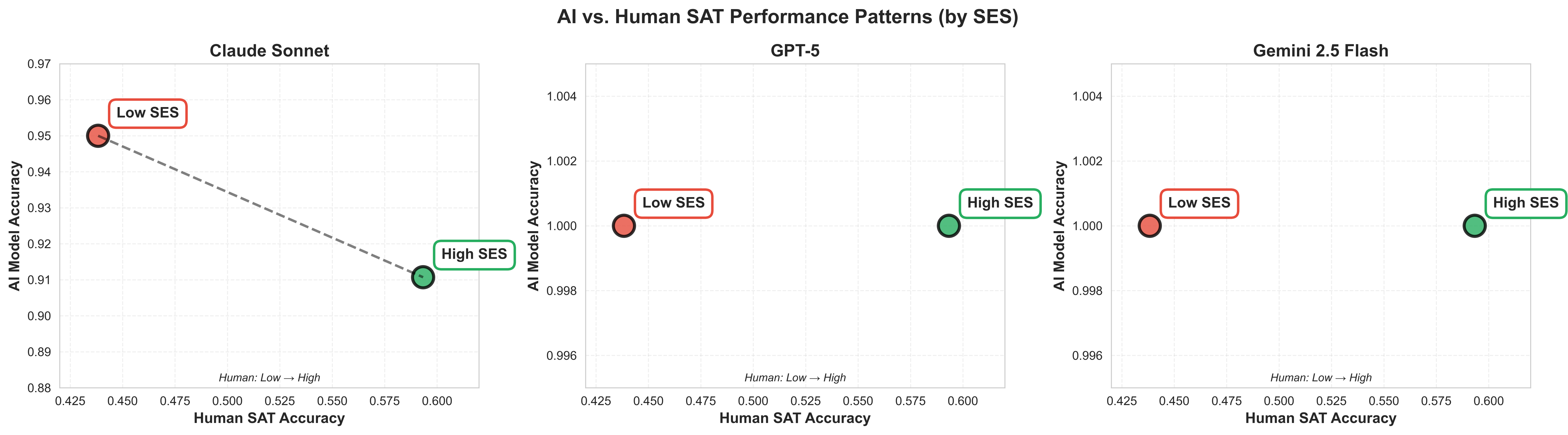}
  \caption{
    \textbf{Alignment between human SAT SES patterns and AI model SES patterns.}
    Scatter points show Low-SES and High-SES performance for humans and each model. A negative slope for Claude Sonnet indicates that the model does not preserve the human-like direction where High-SES students outperform Low-SES students. Whereas GPT-5 and Gemini 2.5 Flash exhibit complete suppression of SES differences. The human accuracy is estimated using College Board reported data \cite{CollegeBoard2007} (see Section \ref{sec4.3.4}). A negative slope for Claude Sonnet indicates inverted alignment (Low-SES personas outperform High-SES personas). GPT-5 and Gemini 2.5 Flash exhibit complete suppression of SES differences (overlapping points at 100\% accuracy), yielding undefined correlation due to zero variance in model SES-based performance.
  }
  \label{fig:human_ai_alignment}
\end{figure}

Finally, we assessed the alignment between human SAT socioeconomic performance patterns and those exhibited by each model. Figure \ref{fig:human_ai_alignment} presents a scatter plot comparing Low-SES and High-SES performance for humans and each model \cite{CollegeBoard2007}. Given that GPT-5 and Gemini 2.5 Flash exhibited complete suppression of SES differences, they offer no meaningful alignment with human patterns (gap $= 0$). Claude Sonnet 4.5, however, showed an inverted SES gap with low-SES outperforming high-SES personas. The presence of a measurable SES gap indicates that Claude Sonnet 4.5 preserves some degree of human-like socioeconomic role-conditioning, albeit in the opposite direction. We explored this phenomenon in the extended replication with 45 personas to assess the robustness of this inversion.

The inversion of the SES performance gap in Claude Sonnet 4.5 suggests that Constitutional AI's contextual reasoning may enable partial retention \cite{Bai2022, Anthropic2025Sonnet45}. However, the directionality misalignment may be due to avoidance of stereotypes or overcompensation during alignment. This highlights the tension between normative alignment goals to avoid harmful typecasting and descriptive fidelity goals.

\subsection{Claude Sonnet 4.5 Extended Replication}
The results from the main experiment indicated that Claude Sonnet 4.5 retained limited but measurable SES-based role-conditioning on the SAT mathematics items. To distinguish this finding from sampling noise and test robustness, we conducted an extended replication with 45 distinct socioeconomic personas (15 each low, middle, high SES) across the same three testing scenarios. This expanded replication included a set of 45 personas over 54 SAT items resulting in 7290 unique SAT responses and 720 preference responses. This increase in statistical power enables detection of small to moderate effects and provides stable estimates of Claude's fidelity. All analyses reported here use deterministic decoding ($T=0.0$); the pertained figures and tables are provided in the Supplementary Materials. The replication results are briefly described in this section.

The replication with 45 personas confirmed the presence of SES-based differences in SAT mathematics accuracy. One way ANOVA showed significant differences between low, middle and high SES agents across all three scenarios with $p<0.05$. Effect sizes were noted to be small to moderate with $\eta^2=0.151$ for optimal, $\eta^2=0.186$ for moderate stress and $\eta^2=0.153$ for challenging scenarios. The directional pattern of performance inversion observed in the main experiment was also replicated, with low-SES personas outperforming high-SES personas across all scenarios. Across the preference tasks, the replication also confirmed robust SES-based variation (Kruskal-Wallis or Chi-squared tests) for nearly all items consistent with the main experiment. The increased power from the larger persona set further revealed statistically significant SES differences in previously non-significant items such as student loans, geographic mobility and emergency savings. The rent vs.\ buy and work flexibility items however remained statistically non-significant ($p>0.05$). The persisting SES-based inversion in SAT performance is likely a property of Claude Sonnet 4.5's alignment mechanisms. The model produces directionally stable but emperically misaligned trait expression under cognitive load.

The clustering analysis analyzed 6947 accurate responses across all 45 agents over the expanded 54 SAT math items and confirmed the limited but measurable SES-based t-SNE clustering observed in the main experiment. While the silhouette score remained near-zero (-0.0025), indicating low middle and high SES are not separable given they answer the same items correctly, hence the reasoning patterns are similar. However, PERMANOVA indicated statistically significant SES-based clustering with $p<0.001$ and $R^2=0.0013$ (see Figure \ref{fig:claude_tsne_ses_clustering}). This suggests the SES while not strongly separable, still exerts a minute influence on the reasoning embeddings. Linguistic analysis of the reasoning outputs further corroborated the main experiment findings. Low-SES personas exhibited longer average sentence lengths (74 vs. 69 words for high-SES, $d=0.24$), while High-SES agents relied on hedging (e.g., "it seems", "probably"), verification steps and mathematical vocabulary more frequently ($|d|<0.2, p < 0.001$). These reflect small structural adjustments in the reasoning but not large scale changes in problem-solving strategies from Figure \ref{fig:claude_reasoning_extended}. The expanded sample reveals that Claude operates primarily through stylistic and structural differences rather than different reasoning approaches. These are quantitative elaborations rather than qualitatively distinct strategies which is consistent with the hypothesis that models cannot reason as constrained personas but rather simulate surface traits.

Finally, the human-AI alignment analysis reaffirmed the inversion of the SES performance gap observed in the main experiment. Figure \ref{fig:ai_human_perf} presents the scatter plot comparing Low-SES and High-SES performance for humans and Claude Sonnet 4.5 across the expanded persona set. The negative slope persisted with low-SES personas outperforming high-SES personas on average across all three scenarios. This inversion suggests that while Claude Sonnet 4.5 preserves some degree of socioeconomic role-conditioning, it does not align with human-like patterns where higher SES typically confers advantages in standardized testing contexts.

\section{Discussion}\label{sec3}
This study explored whether frontier LLMs could maintain distinct socioeconomic role-conditioning while performing cognitively demanding tasks. Across three state-of-the-art models, we observed varying degrees of contextual collapse during SAT mathematics reasoning, with GPT-5 exhibiting complete collapse, Gemini 2.5 Flash showing partial collapse, and Claude Sonnet 4.5 retaining limited role-specific variation. Notably, all models demonstrated distinct role-conditioned affective preferences when cognitive constraints were relaxed. This reveals that while LLMs can achieve distributional fidelity at the population level, they fail to sustain contextual selves under reasoning pressures.

\subsection{Mechanisms of Contextual Collapse}
There are three potential mechanisms that may underlie the observed contextual collapse in role-conditioned LLMs. First, \textit{optimization pressures} result in models choosing to converge upon a singular objective of solving the task directly conflicting with maintaining diverse role-conditioned reasoning patterns. GPT-5 and Gemini 2.5 Flash both exhibited this pattern, where the drive to maximize accuracy led to homogenized responses ($p = 1.000, R^2 = 0.0004$ and $p=0.120, R^2=0.002$). This is despite the robust SES-based variation observed in the preference tasks. Second, \textit{distributional compression} arising from compressed representations of the data distributions. The models have learned the $P(\text{Correct Response} | \text{Pre-Training and Post-Training Alignment})$ while the $P(\text{Correct Response}|\text{Pre-training, Post-training and low-SES student with resource constraints})$ is not sufficiently strong to dominate the output. This results in default solver approach to cognitive tasks. Third, \textit{contextual essentialism} prevents internalization of persona constraints but essentially remain as static prompts. The additional context results in dynamic failure during reasoning, while static prompts have a stronger signal present in the model's innards. This is manifested in Claude's linguistic variations but having identical problem solving strategies. These mechanisms likely interact when distributional compression is strongly felt during optimization pressures, leaving superficial representation of personas through contextual essentialism. This explains why task type determines collapse and having richer training signal is necessary for sustaining role-conditioning.

\subsection{Task-Dependent Trait Expression}
The stark contrast in the results show when LLMs should be deployed as social agents. When cognitive demands are low and the task permits subjective expression, models can exhibit robust role-conditioned variation. LLMs can express differentiated patterns in risk tolerance, time preference and consumption decisions because of multiple valid responses, rich signal in the training data, and lower optimization pressures. When these hold, LLMs may be used as synthetic participants. However, education simulations of different socioeconomic groups, cognitive task simulations or high-stakes decision simulations are likely to induce contextual collapse. The task-dependent trait means reseachers cannot use validation in one domain as a sufficient condition for deployment in another. Each deployment may require a domain-specific validation to ensure role-conditioning fidelity increasing burden on researchers.

\subsection{Prior Work}
Recent work has highlighted that LLMs can be effective social agents and replicate human-like aggregate behaviors in diverse social science tasks \cite{Horton2023, Manning2025, Hu2025}. However, as Wang et al. (2025) compellingly demonstrated, LLMs systematically "misportray and flatten" demographic identities when used as synthetic participants, producing homogenized representations that obscure within-group heterogeneity \cite{Wang2025}. Their analysis of GPT-3.5 and GPT-4 across survey and annotation tasks revealed that models both shift group means away from empirical values and reduce variance within groups, attributing these failures to training data biases and identity prompts that essentialize complex positionalities.

Our work reveals that these problems persist and intensify under cognitive load. While Wang et al. examined static, low-cognitive-demand tasks, we demonstrate that identity representations further deteriorate when models face sustained reasoning pressures. Testing 2025 frontier models (GPT-5, Claude Sonnet 4.5, Gemini 2.5 Flash), we find that GPT-5's complete contextual collapse (PERMANOVA $p = 1.000, R^2 = 0.0004$) represents a more extreme manifestation of Wang et al.'s concerns. The likelihood-based optimization objectives used in current LLM pre-training strategies compress heterogeneity in role-conditioned reasoning, extending Wang et al.'s concerns about demographic misportrayal \cite{Wang2025}. Instruction tuning in post-training encourages helpfulness and truth-seeking, which while beneficial for general alignment, inherently conflicts with persona prompts that ask models to adopt less probable reasoning patterns based on socioeconomic constraints \cite{Sheng2021}. Kim et al. (2024) show that when LLMs are asked to both reason and stay in character, deeper reasoning often leads to abandoning the persona entirely \cite{Kim2024}. The SAT results align with this pattern: as there is only one correct answer to optimize towards, models collapse into a singular identity that maximizes accuracy. The post-training, while offering strong alignment for correctness, compromises the model's ability to sustain diverse contextual selves. The task-dependent trait expression we document being robust on preferences, collapsed on reasoning. This appears to be a fundamental property of current alignment paradigms rather than an artifact of specific demographic attributes with huge implications for suvery based studies and their attention checks.

Claude Sonnet 4.5's partial resistance to collapse while maintaining statistically significant SES-based clustering (PERMANOVA $p < 0.001, R^2 = 0.0043$) and systematic accuracy differences ($\eta^2 = 0.15-0.19$), suggests that alignment methodology influences persona fidelity. However, the inversion of the SES-performance relationship (low-SES personas outperforming high-SES personas) reveals a fundamental tension we term the alignment-fidelity tradeoff. Constitutional AI trains models to consider contextual principles including avoiding stereotyping and being aware of biases. This explicit model training to avoid all toxicity and harmful sterotyping could be a reason for its inversion of performance under low-SES conditioning \cite{Bai2022, Askell2021}. When encountering a 'struggling rural student' persona, these principles may prevent outputs that could be interpreted as stereotype-consistent. The result: Claude maintains that SES matters while inverting how it matters. This represents success at the normative goal (avoiding stereotype-reinforcing outputs) while creating descriptive misalignment with empirical populations where SES genuinely predicts performance due to systemic inequalities in educational access \cite{CollegeBoard2007}.

This finding has important implications for Manning and Horton's (2025) ``General Social Agents", theory-grounded frameworks for LLM agents that generate human-consistent choice distributions \cite{Manning2025}. Our results suggest that even sophisticated alignment methods cannot fully resolve tensions between ethical alignment (not reproducing harmful patterns) and empirical fidelity (accurately simulating populations shaped by structural inequalities). Models may achieve distributional fidelity in aggregate while systematically misrepresenting the mechanisms generating those distributions.

In addition, Stade et al. (2024) argue LLM-driven tools in behavioral healthcare must be evaluated not only on average performance but also on how failure modes map onto human diversity \cite{Stade2024}. Similarly, Gao et al. (2024) emphasize that LLM-powered agent simulations raise concerns about calibration, heterogeneity and interpretability \cite{Gao2024}. Further, Huang et al. (2024) and Zhu et al. (2025) demonstrate that while models may infer human personality from real-world interviews, correlations with validated Big Five scores remain modest \cite{Huang2024, Zhu2025}. Our preference task results both support and complicate these findings: models successfully differentiate SES-based preferences in domains like risk tolerance, time preference, college choice and career priorities, suggesting that SES sentiments regarding these attitudes are deeply embedded in training data. However, the preference task also illuminates potential flattening in statistically insignificant items where models failed to exhibit SES-based variation. Items such as student loans, retirement planning, networking, insurance and renting vs. buying may reflect domains where models' training data did not sufficiently capture nuanced socioeconomic differences, leading to homogenized responses. This is likely due to generic helpful patterns and lack of strong contextual priors in these domains. Narrative signals on burdensome debt, healthcare access, and housing affordability are often complex and multifaceted, and models may default to neutral or average responses in the absence of strong role-conditioning signals.

\subsection{Implications for Survey Research}
Beyond academic simulations, our findings have urgent implications for real-world survey research. The "two-faced" nature of current LLMs, maintaining plausible demographic variation on preference items while collapsing under cognitive load, creates a vulnerability in survey quality control. Traditional attention checks and consistency tests may fail to detect LLM respondents because these models excel at expressing subjectively plausible preferences (Figure \ref{fig:preference_heatmaps}: average effect size $d$=0.55). A malicious actor could deploy persona-conditioned LLMs to systematically bias survey results on opinion polls, consumer preferences, or even election polling, as models demonstrate robust role-conditioned variation precisely on the types of attitudinal questions common in surveys. However, our SAT results suggest a detection strategy: incorporating cognitively demanding tasks that require sustained reasoning under role-specific constraints. While models can express "I prefer candidate X because..." (low cognitive load), they cannot maintain persona-appropriate reasoning patterns when solving multi-step problems that human respondents from different backgrounds would approach differently. The contextual collapse we observe under cognitive load (PERMANOVA $p=1.000$ for GPT-5, $p=0.120$ for Gemini) indicates that deterministic optimization tasks may serve as effective LLM detection mechanisms.

Thus, the following recommendations are proposed for survey researchers concerned about LLM contamination:
\begin{itemize}
  \item \textbf{Incorporate Cognitive Load Tasks:} Embed tasks requiring multi-step reasoning or problem-solving that are sensitive to demographic variation. LLMs are likely to collapse under these conditions, revealing their synthetic nature.
  \item \textbf{Response Pattern Analysis:} Analyze response patterns for homogeneity across demographic groups. Excessive uniformity may indicate LLM involvement.
  \item \textbf{Timing Analysis:} As LLMs respond rapidly, unusually fast completion times may signal synthetic respondents for cognitively demanding sections.
  \item \textbf{Linguistic Fingerprint:} Use structural markers indicated in our linguistic analysis (see Section \ref{sec4})
\end{itemize}

The fact that Claude Sonnet 4.5's low-SES personas systematically outperformed high-SES personas on SAT items (inverting real human patterns; Figure \ref{fig:human_ai_alignment}) suggests that even the most sophisticated alignment approaches have not solved the challenge of authentic constraint-based reasoning. Until models can sustain realistic cognitive limitations across task types, their use in survey research poses risks that extend beyond academic validity to public discourse integrity.

\subsection{Limitations}
Several limitations of our study should be considered. First, our main analyses use deterministic decoding ($T=0.0$), which likely magnifies convergence; robustness checks at $T=0.6$ (reported in the Supplementary) suggest that stochasticity did not significantly alter the results, the overall pattern of contextual collapse under SAT reasoning remains. Second, our agents are memoryless, each response is conditioned on a persona description and scenario, but not on a persistent history. Frameworks that endow agents with long-term memory or state (e.g., persisting agent architectures) might support more stable role representations over time. Third, we focus on SES-based personas and deliberately avoid protected demographic attributes to reduce the risk of replicating harmful stereotypes, as highlighted by Wang et al.~\cite{Wang2025}. This choice narrows the scope of our conclusions but aligns with ongoing discussions about ethical experimentation with demographic prompts. Finally, we study a relatively narrow domain of analytic reasoning (SAT mathematics). Persona fidelity might behave differently in domains where social identity is more central, such as moral dilemmas, political opinion, or interpersonal negotiation. These limitations provide meaningful avenues for future research.

\section{Methods}\label{sec4}
In this section, we detail the experimental design, including personas, task construction, model configurations, and evaluation metrics. First, the models chosen are the current frontier LLMs as of late 2025: GPT-5, Claude Sonnet 4.5, and Gemini 2.5 Flash \cite{OpenAI2025GPT5, Anthropic2025Sonnet45, GoogleDeepMind2025Gemini2.5Flash}. GPT-5's accuracy-optimized RLHF training was hypothesized to drive contextual collapse, as optimization pressure toward singular correct answers may override persona constraints particularly concerning given GPT-5's market share making it the most likely vector for both legitimate simulations and potential survey contamination \cite{TullyEtAl2025,Westwood2025}. Each model was accessed via the ExpectedParrot API which provides a unified interface for querying multiple LLMs along with agent creation and scenario management \cite{Horton2024EDSL}. The chosen models present an opportunity to evaluate the latest advancements in LLM capabilities and alignment strategies; they also dominate the market share among enterprise LLMs \cite{TullyEtAl2025}. As role-conditioning becomes more prevalent in social simulations (see Section \ref{sec1}), understanding how these frontier models mask their cognitive capabilities under persona constraints becomes important. While all three models have been tuned to be maximally truth seeking and helpful, their human counterparts are often boundedly rational \cite{Simon1955, Simon1957}, influenced by their social identities \cite{TajfelTurner1979, AkerlofKranton2000} and fundamentally fallible \cite{TverskyKahneman1974, Kahneman2011}.

\subsection{Experimental Overview}
The experiment was conducted in two phases: a main experiment evaluating all three models across 15 socioeconomic personas and a replication focusing on Claude Sonnet 4.5 with an expanded set of 45 personas given its superior role-conditioning fidelity observed in the main experiment (see Section \ref{sec2}). Each model was evaluated on two task types: SAT mathematics items and affective preference tasks (see Table \ref{tab:experiment_design}). The SAT items were drawn randomly from a collection of SAT math problems from official tests that appeared in 2007, the corresponding SAT reports included student performance data across socioeconomic strata \cite{CollegeBoard2007}. Additionally, the scenario was manipulated to simulate varying cognitive stress levels: optimal (no stress), moderate stress (time pressure), and challenging (time pressure with distractions). Each persona was evaluated across all three scenarios in the main and replication experiments. Over the entire experiment, each model was prompted to simulate 15 agents across 3 scenarios providing a total of 3780 unique responses for SAT items and 720 for preference tasks. For Claude Sonnet 4.5 in the replication, 45 agents across 3 scenarios provided 7290 unique SAT responses for the SAT items and 720 preference responses.

The affective preference tasks were adapted from established economic and social psychology literature assessing subjective preferences across domains such as risk tolerance, time preference, social trust, and moral dilemmas. Each task was designed to elicit responses that could vary meaningfully based on the assigned socioeconomic persona. Table \ref{tab:preference_tasks} summarizes the preference task domains and representative literature from which the questions were adapted. These questions were prompted in both main and replication experiments without scenario variations, as they were assumed to be stable across stress conditions. We used deterministic ($T=0.0$) and stochastic ($T=0.6$) decoding for the main experiment to assess the impact of response variability on role-conditioning fidelity, while only deterministic decoding was used in the replication to focus on high-fidelity role enactment. The main text presents the results only from the deterministic decoding condition; results from the stochastic decoding are provided in the Supplementary Materials as they exhibited near identical patterns.

\subsection{Persona Construction}
Fifteen unique socioeconomic personas were constructed to represent a diverse range of backgrounds across income, education, and occupation dimensions. Each persona was defined by a detailed profile including a name, grade level, parental income bracket, parental education level, parental occupation, testing experience, testing preparation, geographic location of school and personal hobbies. Given that Wang et al. (2025) highlighted the risks of demographic misportrayal, we deliberately avoided using identities but allowed the model to infer the persona through the socioeconomic characterization rather than explicit demographic markers \cite{Wang2025}. These profiles were binned into low (5), middle (4), and high (6) socioeconomic status (SES) categories based on parental income and education levels. Personas span a full range of socioeconomic and educational contexts, including variation in parental education, school resources, and access to test preparation. Each agent served as a distinct synthetic participant in the standardized SAT reasoning evaluation. The replication experiment expanded this set to 45 personas (15 each low, middle, high SES) by varying the attributes more granularly across the same dimensions. The full list of personas used in replication experiment are provided in the Supplementary Materials. The 15 personas used in the main experiment are provided in Table \ref{tab:personas}.

These personas were instantiated using the \texttt{edsl.Agent} framework from the ExpectedParrot API, which allows for structured agent definitions and scenario management \cite{Horton2024EDSL}. EDSL enables automatic system prompt generation based on the defined persona attributes, ensuring consistent prompt construction across models and tasks. Each agent was prompted with a standardized SAT reasoning prompt or preference task prompt, embedding their socioeconomic profile to guide their responses.

\subsection{Analytical Framework}
\subsubsection{SAT and Preference Task Evaluation}
Having collected the responses from the models across agents, tasks and scenarios, the analyses were conducted using Python 3.11 with standard scientific libraries (scipy, numpy, pandas, scikit-learn, sentence-transformers and seaborn). One-way ANOVA tests were used to assess differences in mean performances across the three SES groups for SAT items for each model. Effect sizes were computed to be the proportion of total variance in SAT accuracy attributable to the difference between SES groups ($\eta^2 = \frac{\text{Between-Group Sum of Squares}}{\text{Total Sum of Squares}}$). The size of the effects were interpreted using Cohen's (1988) guidelines for $\eta^2$ where small = 0.01, medium = 0.06 and large = 0.14 \cite{Cohen1988}.

For the preference task, the 16 items were recoded into ordinal or categorical variables based on their underlying economic constructs. Ordinal items included risk, time, education and consumption preferences and were coded on a four-point ordered scale where higher values indicated greater risk tolerance, patience, education prioritization and consumption preference respectively. The other items such as career priorities, networking style and health-insurance choice were retained as categorical variables. The ordinal coded variables were evaluated using Kruskal-Wallis H test, a rank based non-parametric analogue of the one-way ANOVA test for the 3 SES groups. Effect sizes were computed using epsilon squared ($\epsilon^2 = \frac{H}{n - k}$) where $H$ is the Kruskal-Wallis H statistic, $k$ is the number of groups and $n$ is the total sample size. The categorical variables on the other hand were tested using the Chi-squared test of independence with Cramer's V ($V = \sqrt{\frac{\chi^2/n}{\min(k-1, r-1)}}$) as the effect size measure where $k$ and $r$ are the number of columns (unique responses available per question item) and rows (3 SES classes) respectively. We interpret the effect sizes using Cohen's (1988) guidelines for $\epsilon^2$ and Cramer's V where small = 0.1, medium = 0.3 and large = 0.5 \cite{Cohen1988}.

\subsubsection{Semantic Embedding Analysis}
The semantic embedding analysis was conducted to assess the overall role-conditioning fidelity across the full response set for each model. Using the \texttt{sentence-transformers/all-MiniLM-L6-v2}, a widely used sentence embedding model, each reasoning passage response was converted into a 384-dimensional vector representation \cite{Reimers2019SBERT}. The embeddings were optimized for semantic similarity tasks, providing an effective balance between lingusitic coverage and efficiency. The reasoning passages were extracted from the model responses marked correct based on the official SAT answer key and encoded into embeddings with batch size of 32 and L2-normalized to unit length. For each response $i$, the model produced a vector representation $\mathbf{e}_i \in \mathbb{R}^{384}$, such that the cosine-distance between two embeddings $\mathbf{e}_i$ and $\mathbf{e}_j$ is given by:
$$
  \text{d}_{ij}(\mathbf{e}_i, \mathbf{e}_j) = 1 - \frac{\mathbf{e}_i \cdot \mathbf{e}_j}{\|\mathbf{e}_i\| \|\mathbf{e}_j\|}
$$
Where $\cdot$ denotes the dot product and $\|\cdot\|$ is the Euclidean norm. This distance metric captures the semantic dissimilarity between two responses in the embedding space. By focusing on correct responses, we ensured that the analysis centered on the reasoning patterns rather than accuracy differences.

To evaluate whether reasoning represented SES relevant structure, we analyzed each model's embedding space for their cluster separation, statistical significance and visual interpretation. First, we computed the average silhouette score ($S$) for each model's embedding space, defined as:
$$
  S = \frac{1}{N} \sum_{i=1}^{N} \frac{b(i) - a(i)}{\max\{a(i), b(i)\}}
$$
For each data point $i$, where $a(i)$ is the average distance between $i$ and all other points in the same SES cluster, $b(i)$ is the average nearest-cluster distance for point $i$, and $N$ is the total number of points. The silhouette score ranges from -1 to 1, with higher values indicating better-defined clusters. A score close to 0 suggests overlapping clusters, while negative values indicate potential misclassification \cite{Rousseeuw1987}. The statistical testing of group separation was conducted using Permutational Multivariate Analysis of Variance (PERMANOVA) on the pairwise cosine-distance matrix. PERMANOVA which is a non-parametric method ideal for testing group differences in a high-dimensional space based on a distance metric tests whether the centroids of different groups (SES classes) are significantly different in the embedding space. The pseudo-F test statistic is computed as:
$$
  F = \frac{\text{SS}_{\text{between}} / (k - 1)}{\text{SS}_{\text{within}} / (N - k)}
$$
Where $\text{SS}_{\text{between}}$ is the sum of squares between groups, $\text{SS}_{\text{within}}$ is the sum of squares within groups, $k$ is the number of groups, and $N$ is the total number of observations. The significance of the $F$ statistic is assessed through permutation testing, where group labels are randomly shuffled 999 times to generate a null distribution \cite{Anderson2017}. After which, we computed the effect size as $R^2 = \frac{\text{SS}_{\text{between}}}{\text{SS}_{\text{total}}}$, representing the proportion of variance explained by group differences. Finally, for visual interpretation, we employed t-SNE to reduce the high-dimensional embedding space to two dimensions while preserving local structure with cosine distance, perplexity of 30 and learning rate of 300 over 1000 iterations \cite{vanDerMaaten2008tSNE}. This allowed us to visually identify whether distinct SES groups occupy statistically distinguishable regions of reasoning space.

\subsubsection{Linguistic Feature Analysis}
Further, we conducted lingustic and stylistic feature analysis to understand how role-conditioning influenced the models' language use during reasoning. Using measurable markers of structure, meta-cognition and confidence, we extracted multiple features from the reasoning tokens of each model's responses. All linguistic features were extracted programmatically using Python 3.12 with a rule-based pattern matching and regular expressions approach. The reasoning tokens were classified into the following categories:
\begin{itemize}
  \item \textbf{Length and fluency:} Total character count, words, sentences and average word and sentence length.
  \item \textbf{Structural markers:} Presence of question restatement, stepwise organization and verification phrases (e.g., ``therefore,'' ``thus''). The regex: [-*•]|\textbackslash{}d+\textbackslash{}. was used to identify stepwise organization, bullets or numbered lists.
  \item \textbf{Lingusitic style indicators:} Use of hedging language or uncertainty expressions (e.g., ``might,'' ``could be'', ``I think''), meta commentary on reasoning process (e.g., ``let me see,'' ``I need to," ``we can") and pronoun counts (first-person vs. plural). First-person singular counted using regex \textbackslash{}b(i|me|my|mine)\textbackslash{}b; first-person plural using \textbackslash{}b(we|us|our|ours)\textbackslash{}b (case-insensitive).
  \item \textbf{Lexical diversity:} Proportion of unique tokens (vocabulary richness) and frequency of technical math terms (e.g., ``equation,'' ``variable,'' ``substitute,'' ``solve''). Counted domain-specific terms including "equation", "solve", "substitute", "simplify", "calculate", "compute", "determine", "formula", "variable", "constant", "coefficient", "expression", "function", "derivative", "integral" (15 terms total).
  \item \textbf{Confidence and emphasis:} Count of exclamation marks and step by step labeling phrases (e.g., ? symbol or the ! mark).
\end{itemize}
These features were normalized per 100 words to control for the variation in response length. For each model, the low vs. high SES groups were compared using t-tests and effect sizes computed using Cohen's d \cite{Cohen1988}. This captured the interpretable linguistic correlations of the semantic clustering patterns observed in the embedding analysis. complete feature extraction code is provided in Supplementary Materials (see Table~\ref{tab:linguistic_features} for complete feature definitions).

\subsubsection{Alignment with Human Data}\label{sec4.3.4}
To verify whether the models' role-conditioned responses aligned with known human socioeconomic conditioned outcomes, we compared the model's low and high-SES accuracy gap to that observed among human SAT test-takers from the College Board's 2007 report \cite{CollegeBoard2007}. By restricting low and high SES definitions to match those used in the report (parental income below \$30,000 and above \$100,000 respectively), we computed an estimate for accuracy for human test-takers in each group using the following formula:
$$
  \text{Accuracy} = \frac{\text{Mean Score} - \text{Min Score}}{\text{Max Score} - \text{Min Score}} = \frac{\text{Mean Score} - 200}{800 - 200}
$$
The resulting accuracy gap using the average scores for low SES (mean = 463) and high SES (mean = 556) for the math section was 15.5 percentage points. Similarly, we computed the accuracy gap for each model as $\text{Accuracy}_{\text{high SES}} - \text{Accuracy}_{\text{low SES}}$ to assess how closely the models' role-conditioned performance mirrored human socioeconomic disparities. To assess the directional alignment of the models we computed the Pearson correlation between the ordered human and model accuracies. Given that the comparison is across low and high SES groups, the correlation is trivially either 1 (aligned) or -1 (misaligned). This analysis provided a direct test of whether AI alignment strategies replicate, suppress or invert empirical human socioeconomic patterns in high-stakes reasoning tasks. For models exhibiting complete SES suppression (GPT-5 and Gemini 2.5 Flash, both showing 100\% accuracy across all SES groups), we assigned a "suppression" category rather than computing correlation, as the zero-variance condition renders Pearson's $r$ undefined.

\section{Data and Code Availability}
All the replication requirements including code to generate the results presented in the analysis are available at \url{https://github.com/krishnaveti/twofaced_socialagents}.

\begin{appendix}

  \section{Extended Results and Tables}\label{sec7}
  This section contains extended figures and tables referenced in the main text.

  \begin{table}[ht!]
    \centering
    \caption{\textbf{Preference task domains and representative literature:}
      Each of the sixteen preference questions was designed to reflect a canonical construct in behavioral and labor economics.
      Domains are grouped by citation set, with two items per construct unless otherwise noted.}
    \label{tab:preference_tasks}
    \renewcommand{\arraystretch}{1.25}
    \small
    \begin{tabular}{p{4.6cm}p{7.0cm}p{2.0cm}}
      \toprule
      \textbf{Domain (Item Count)}          & \textbf{Question Focus}                                               & \textbf{Representative Sources}                  \\
      \midrule
      \textbf{Risk Preferences (2)}         & Financial risk tolerance; job security vs.\ income variability        & \cite{holt_laury_2002, dohmen_2011}              \\[3pt]

      \textbf{Time Preferences (2)}         & Temporal discounting; education investment horizon                    & \cite{frederick_2002, chabris_2008}              \\[3pt]

      \textbf{College Choice \& Debt (2)}   & College selection by cost vs.\ prestige; student loan comfort         & \cite{avery_hoxby_2004, lovenheim_reynolds_2013} \\[3pt]

      \textbf{Labor Market (2)}             & Career priorities (security, salary, passion); work--life flexibility & \cite{wiswall_zafar_2018, mas_pallais_2017}      \\[3pt]

      \textbf{Consumption \& Lifestyle (2)} & Windfall spending; car purchase decisions                             & \cite{charles_2009}                              \\[3pt]

      \textbf{Social Capital (2)}           & Geographic mobility for career; networking approach                   & \cite{granovetter_1973, chetty_2022}             \\[3pt]

      \textbf{Financial Planning (2)}       & Retirement planning engagement; emergency savings targets             & \cite{lusardi_mitchell_2014}                     \\[3pt]

      \textbf{Health \& Insurance (1)}      & Health insurance selection priorities                                 & \cite{finkelstein_2019}                          \\[3pt]

      \textbf{Housing Preferences (1)}      & Rent vs.\ buy preferences                                             & \cite{lusardi_mitchell_2014}                     \\
      \bottomrule
    \end{tabular}

    \vspace{0.4em}
    \footnotesize
    \textit{Note.} The task also included one validation question prompting each agent to recall its assigned persona name to ensure consistent identity retention throughout the assessment.
  \end{table}

  \begin{table}
    \centering
    \caption{\textbf{Socioeconomic and educational personas used in the main experiment.}
      Fifteen distinct agents were defined to capture variation in socioeconomic background, educational context, and test preparation access.}.
    \label{tab:personas}
    \vspace{0.5em}
    \renewcommand{\arraystretch}{1.25}
    \small
    \begin{tabular}{p{4.2cm}p{9.5cm}}
      \toprule
      \textbf{Persona Name}                           & \textbf{Summary Traits}                                                                                                                                              \\
      \midrule
      \textbf{struggling\_rural\_student (Emma)}      & 10th-grade student from rural area; struggles with basic algebra; low-income, first-generation college aspirant; works on family farm; unfamiliar with SAT.          \\[3pt]

      \textbf{urban\_underresourced (Marcus)}         & 11th-grader in underfunded urban school; single-parent household; below-average math skills; limited prep; works after school; moderate test anxiety.                \\[3pt]

      \textbf{ell\_student (Sofia)}                   & 9th-grade English-language learner; strong conceptual math but language barriers; working-class immigrant family; high test anxiety due to comprehension challenges. \\[3pt]

      \textbf{distracted\_suburban (Tyler)}           & 10th-grade suburban student; capable but inattentive; middle-class background; minimal prep; low motivation, low anxiety.                                            \\[3pt]

      \textbf{rural\_average (Hannah)}                & 11th-grader from small rural town; working-class; self-studies via Khan Academy; solid fundamentals, unsure about college path.                                      \\[3pt]

      \textbf{typical\_suburban (Ethan)}              & 11th-grader; B-student in math; middle-class two-parent home; took summer SAT prep; moderate anxiety.                                                                \\[3pt]

      \textbf{urban\_motivated (Aisha)}               & 11th-grader in charter school; lower-middle-class immigrant family; strong work ethic; heavy use of free prep; high family expectations.                             \\[3pt]

      \textbf{small\_town\_steady (Noah)}             & 12th-grader; consistent B+ student; middle-class small-business family; limited AP options; low-moderate anxiety.                                                    \\[3pt]

      \textbf{college\_prep\_achiever (Olivia)}       & 11th-grader; excels in math (AP Calculus AB); upper-middle-class; private SAT tutor; multiple enrichment activities.                                                 \\[3pt]

      \textbf{magnet\_school\_star (Jayden)}          & 12th-grader; top performer in selective STEM magnet; advanced curriculum; robotics captain; low test anxiety.                                                        \\[3pt]

      \textbf{motivated\_upward\_mobility (Maya)}     & 12th-grader; first-generation college student from low-income household; highly self-driven; extensive free prep; high anxiety due to scholarship stakes.            \\[3pt]

      \textbf{private\_school\_elite (Alexander)}     & 12th-grader in elite private school; affluent family; extensive tutoring (50+ hours); multiple test sittings; Ivy-targeted preparation.                              \\[3pt]

      \textbf{math\_competition\_specialist (Daniel)} & 11th-grader; AMC/AIME qualifier; STEM-professional parents; minimal formal prep; very low anxiety; motivated by intrinsic math interest.                             \\[3pt]

      \textbf{prodigy\_accelerated (Zoe)}             & 10th-grader taking university real analysis; identified early as gifted; attends college courses; minimal prep needed.                                               \\[3pt]

      \textbf{well\_rounded\_achiever (Liam)}         & 12th-grader excelling across subjects; upper-middle-class family with professional degrees; comprehensive prep; holistic achiever profile.                           \\
      \bottomrule
    \end{tabular}
  \end{table}

  \begin{table}[h]
    \centering
    \caption{Complete Linguistic Feature Definitions and Extraction Methods}
    \label{tab:linguistic_features}
    \small
    \begin{tabular}{p{2.5cm}p{3cm}p{2.5cm}p{6cm}}
      \toprule
      \textbf{Feature Category} & \textbf{Feature Name}          & \textbf{Extraction Method} & \textbf{Search Terms/Patterns}                                                                                                                                                                                         \\
      \midrule

      \multicolumn{4}{l}{\textit{Structural Markers}}                                                                                                                                                                                                                                                                  \\
                                & Question restatement           & Keyword matching           & ``the problem'', ``the question'', ``this asks'', ``this is asking'', ``we're asked'', ``we are asked'', ``we want to find'', ``we need to find''                                                                      \\
      \addlinespace[0.5em]

                                & Stepwise organization          & Keyword + regex            & ``step 1'', ``step 2'', ``first'', ``second'', ``third'', ``finally'' + regex: \texttt{[-*\textbullet{}]|\textbackslash{}d+\textbackslash{}.}                                                                          \\
      \addlinespace[0.5em]

                                & Verification phrases           & Keyword matching           & ``check'', ``verify'', ``confirm'', ``make sure'', ``ensures'', ``correct'', ``therefore'', ``so'', ``thus'', ``hence''                                                                                                \\
      \addlinespace[0.5em]

      \midrule
      \multicolumn{4}{l}{\textit{Linguistic Style Indicators}}                                                                                                                                                                                                                                                         \\
                                & Hedging language               & Keyword matching           & ``maybe'', ``perhaps'', ``possibly'', ``might'', ``could'', ``may'', ``seems'', ``appears'', ``likely'', ``probably''                                                                                                  \\
      \addlinespace[0.5em]

                                & Uncertainty expressions        & Keyword matching           & ``i think'', ``i believe''                                                                                                                                                                                             \\
      \addlinespace[0.5em]

                                & Meta-commentary                & Keyword matching           & ``let me'', ``i need to'', ``i'll'', ``i will'', ``first'', ``then'', ``next'', ``looking at'', ``given'', ``we need'', ``we can'', ``we have'', ``we know'', ``let's'', ``i see'', ``i'm'', ``i am''                  \\
      \addlinespace[0.5em]

                                & First-person singular pronouns & Regex (case-insensitive)   & \texttt{\textbackslash{}b(i|me|my|mine)\textbackslash{}b}                                                                                                                                                              \\
      \addlinespace[0.5em]

                                & First-person plural pronouns   & Regex (case-insensitive)   & \texttt{\textbackslash{}b(we|us|our|ours)\textbackslash{}b}                                                                                                                                                            \\
      \addlinespace[0.5em]

      \midrule
      \multicolumn{4}{l}{\textit{Lexical Diversity}}                                                                                                                                                                                                                                                                   \\
                                & Vocabulary richness            & Type-token ratio           & Unique words / total words (after lowercasing)                                                                                                                                                                         \\
      \addlinespace[0.5em]

                                & Mathematical vocabulary        & Keyword matching           & ``equation'', ``solve'', ``substitute'', ``simplify'', ``calculate'', ``compute'', ``determine'', ``formula'', ``variable'', ``constant'', ``coefficient'', ``expression'', ``function'', ``derivative'', ``integral'' \\
      \addlinespace[0.5em]

      \midrule
      \multicolumn{4}{l}{\textit{Confidence \& Emphasis Markers}}                                                                                                                                                                                                                                                      \\
                                & Exclamation marks              & Direct count               & !                                                                                                                                                                                                                      \\
      \addlinespace[0.5em]

                                & Question marks                 & Direct count               & ?                                                                                                                                                                                                                      \\

      \bottomrule
    \end{tabular}

    \vspace{0.5em}
    \raggedright
    \footnotesize
    \textit{Note:} All text matching was performed case-insensitively on lowercased text. Count-based features were normalized per 100 words using the formula: (feature\_count / word\_count) $\times$ 100. Regex patterns used Python syntax with word boundary anchors (\texttt{\textbackslash{}b}) to match whole words only.
  \end{table}

  \begin{figure}[ht]
    \centering
    \includegraphics[width=0.95\textwidth]{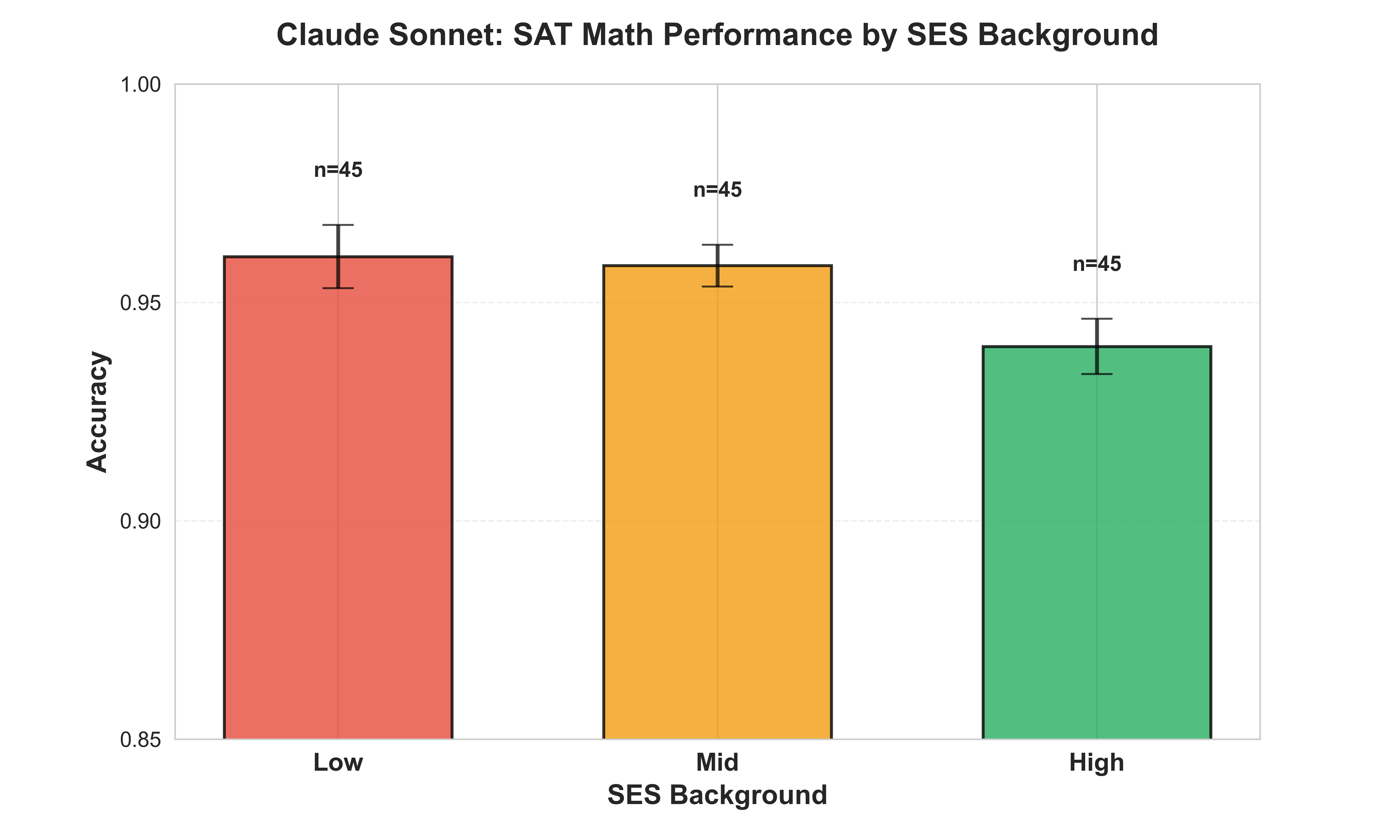}

    \caption{
      \textbf{Extended replication: SES–performance gradient in Claude Sonnet.}
      Results from a larger-scale replication using \emph{N=45} SES-conditioned agents
      (135 SAT observations) confirm a consistent SES-linked accuracy gradient in Claude Sonnet.
      Increased sample size provides substantially greater statistical power, enabling detection
      of moderate effects with stability across random agent initializations.
      The replicated pattern supports the central finding that Claude’s
      Constitutional AI alignment permits systematic socioeconomic trait embodiment
      in mathematical reasoning performance.
    }
    \label{fig:claude_replication_accuracy}
  \end{figure}

  \begin{figure}[ht]
    \centering

    \begin{subfigure}[t]{0.48\textwidth}
      \centering
      \includegraphics[width=\textwidth]{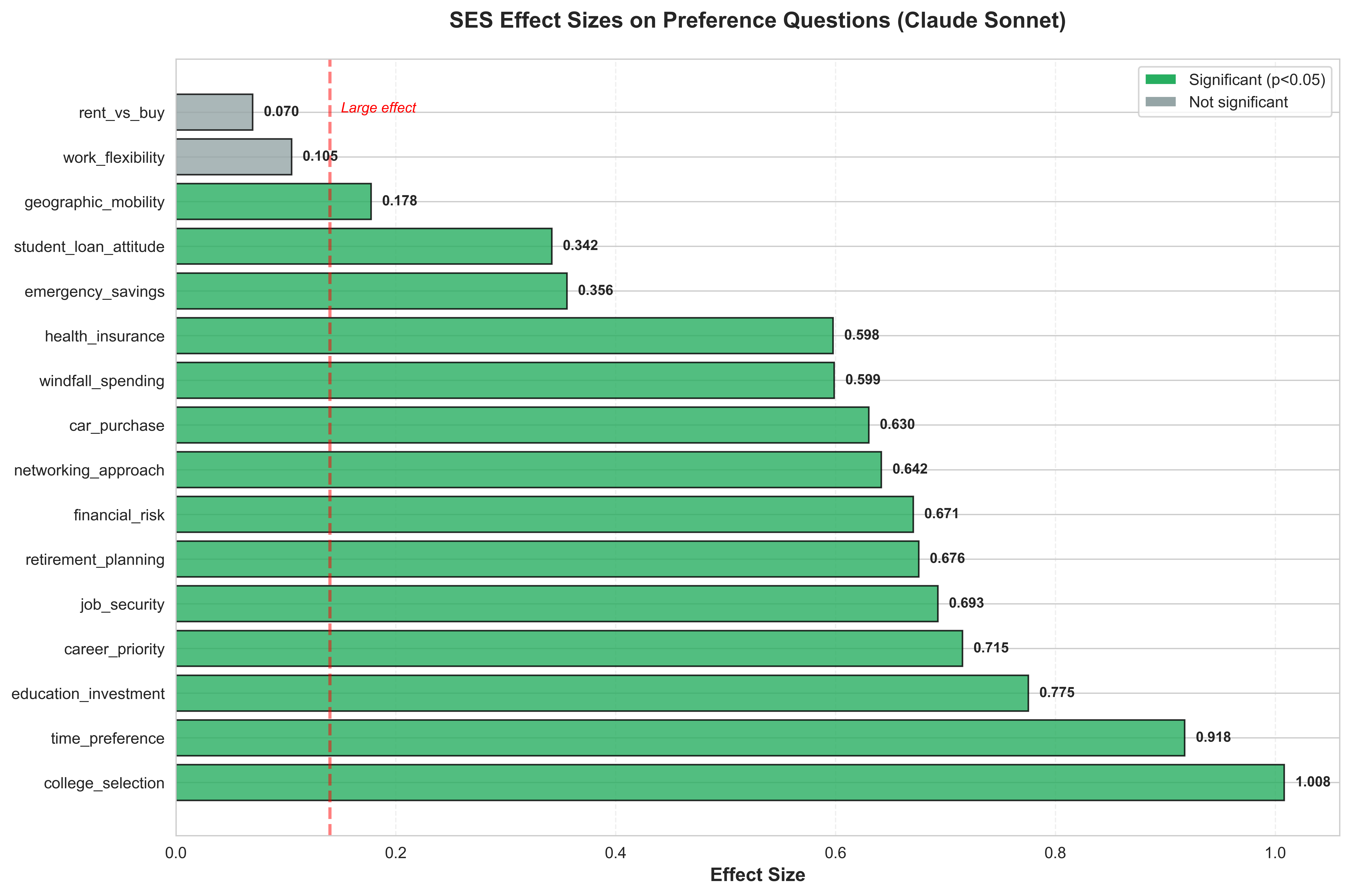}
      \caption*{\textbf{(a)} Effect sizes across 16 preference domains.}
    \end{subfigure}
    \hfill
    \begin{subfigure}[t]{0.48\textwidth}
      \centering
      \includegraphics[width=\textwidth]{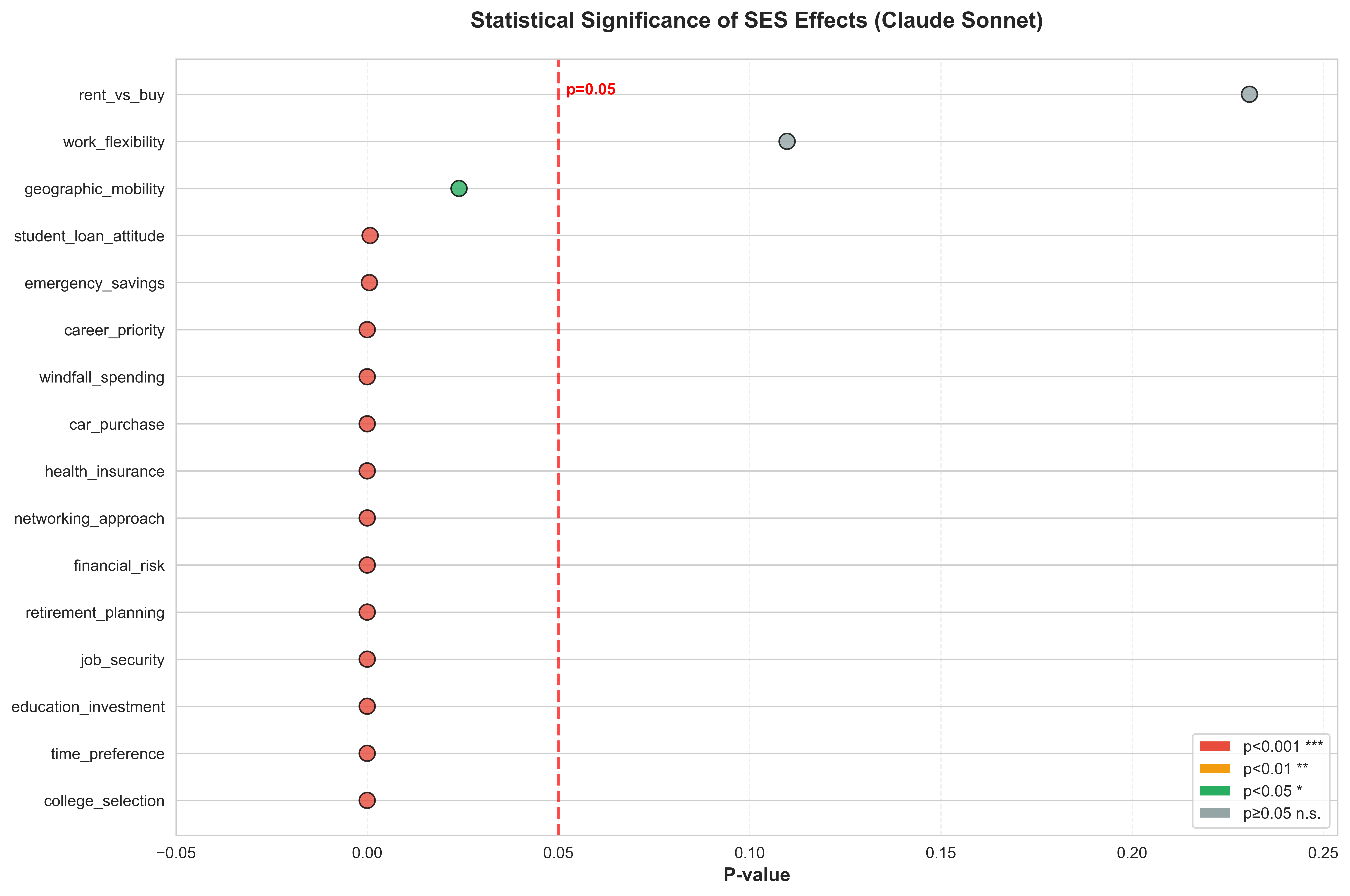}
      \caption*{\textbf{(b)} Statistical significance across domains.}
    \end{subfigure}

    \caption{
      \textbf{Claude Sonnet replication (45 agents): differential SES embodiment in preference judgments.}
      A replication using \emph{N=45} SES-conditioned agents (three times the baseline sample) reveals stable
      cross-domain differences in affective and economic preferences.
      Panel \textbf{(a)} shows effect sizes (\(\epsilon^2\) for ordinal items; Cramer’s \(V\) for
      categorical items). Panel \textbf{(b)} shows corresponding p-values.
      Results demonstrate consistent SES-linked variation in domains such as risk tolerance,
      time preferences, college choice, and consumption decisions, with fewer domains (e.g.,
      work flexibility, rent vs.\ buy) showing no detectable SES effects.
    }
    \label{fig:claude_preference_replication}
  \end{figure}

  \begin{figure}[ht]
    \centering
    \includegraphics[width=0.95\textwidth]{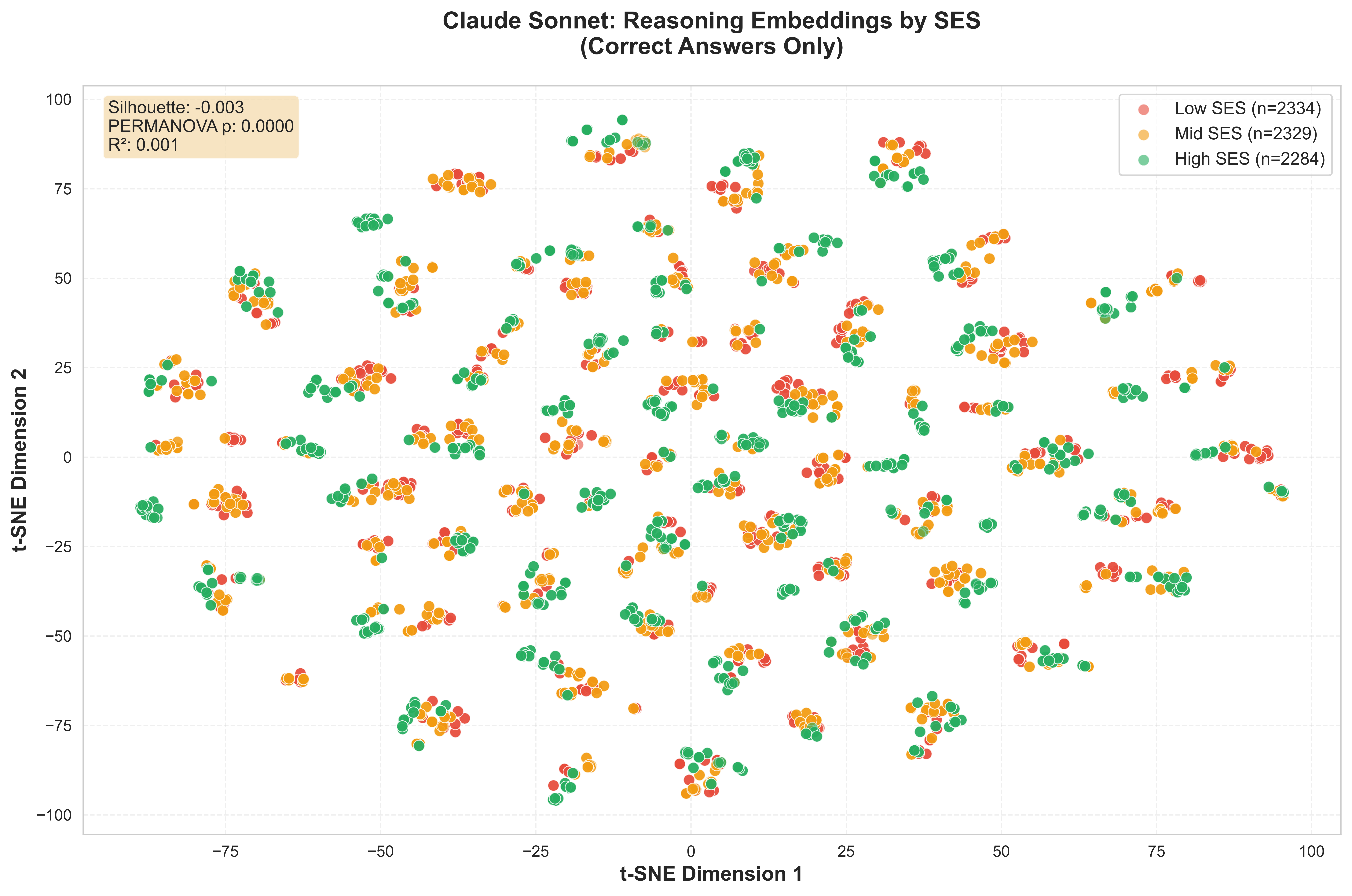}

    \caption{
      \textbf{t-SNE projection of semantic embeddings (Claude Sonnet, extended sample).}
      Two-dimensional t-SNE visualization of 45 agents’ SAT reasoning embeddings,
      colored by socioeconomic status (SES). Despite the large sample, SES groups
      form highly overlapping clusters, indicating minimal separability in the
      underlying embedding space. Consistent with PERMANOVA results (see \ref{sec2}),
      SES explains only a negligible proportion of variance, suggesting that SES-linked
      reasoning differences, remain stylistically subtle rather than structurally distinct.
    }
    \label{fig:claude_tsne_ses_clustering}
  \end{figure}

  \begin{figure}[ht]
    \centering

    \begin{subfigure}[t]{0.48\textwidth}
      \centering
      \includegraphics[width=\textwidth]{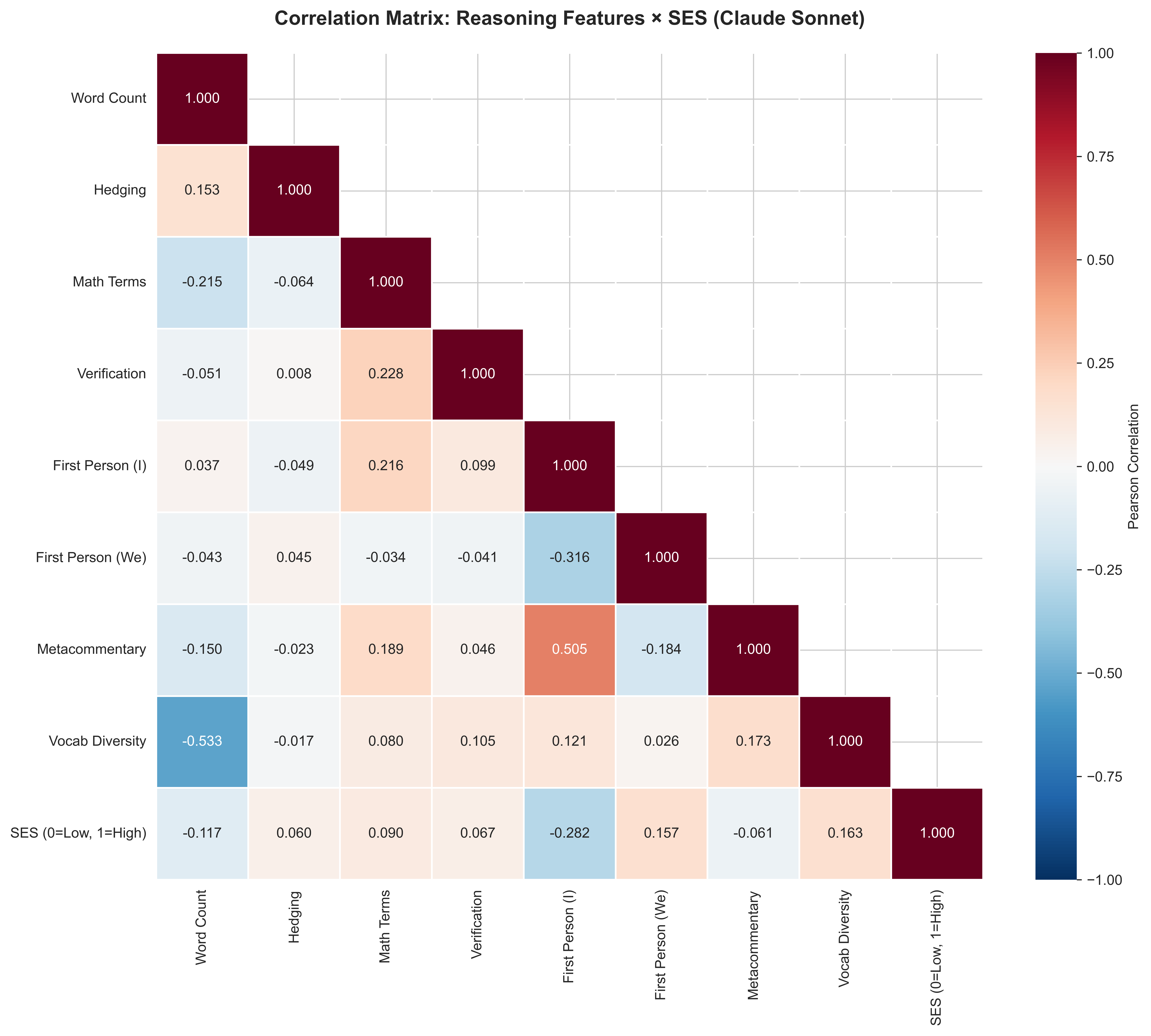}
      \caption*{\textbf{(a)} Correlation structure of linguistic features across SES.}
    \end{subfigure}
    \hfill
    \begin{subfigure}[t]{0.48\textwidth}
      \centering
      \includegraphics[width=\textwidth]{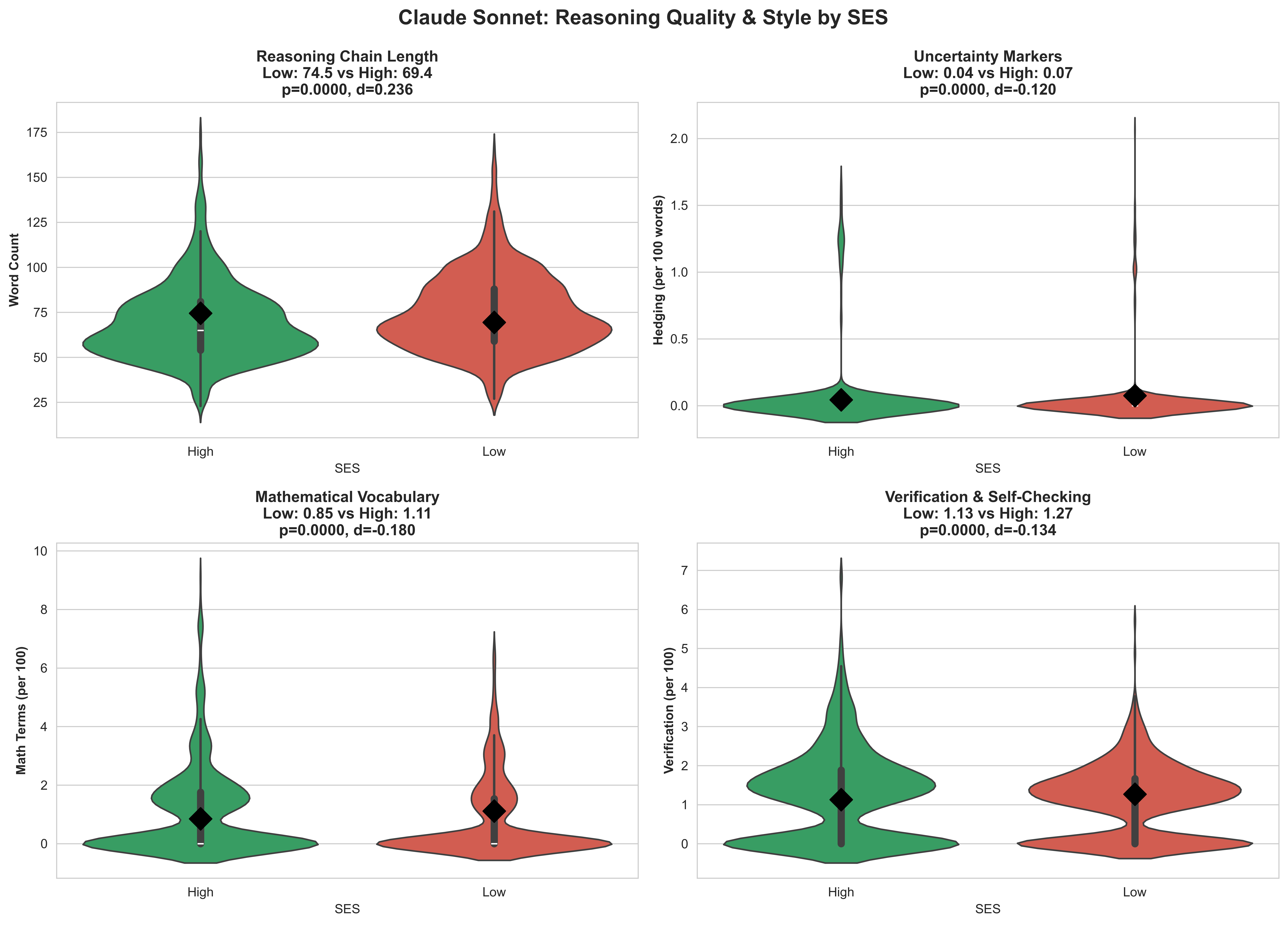}
      \caption*{\textbf{(b)} SES differences in reasoning length, verification, vocabulary, and hedging.}
    \end{subfigure}

    \caption{
      \textbf{Extended reasoning-quality analysis using 45 Claude Sonnet agents.}
      Panel \textbf{(a)} shows the correlation matrix linking SES (Low–High) to linguistic
      and structural reasoning features in correct SAT solutions.
      Panel \textbf{(b)} displays SES-conditioned distributions for key reasoning markers—
      including chain length, mathematical vocabulary, verification phrases, and uncertainty markers.
      The enlarged agent pool (\(N=45\)) confirms consistent SES-conditioned differences in
      reasoning structure and metacognitive expression, reproducing and strengthening
      the patterns observed in the baseline analysis.
    }
    \label{fig:claude_reasoning_extended}
  \end{figure}

  \begin{figure}[ht]
    \centering
    \includegraphics[width=0.95\textwidth]{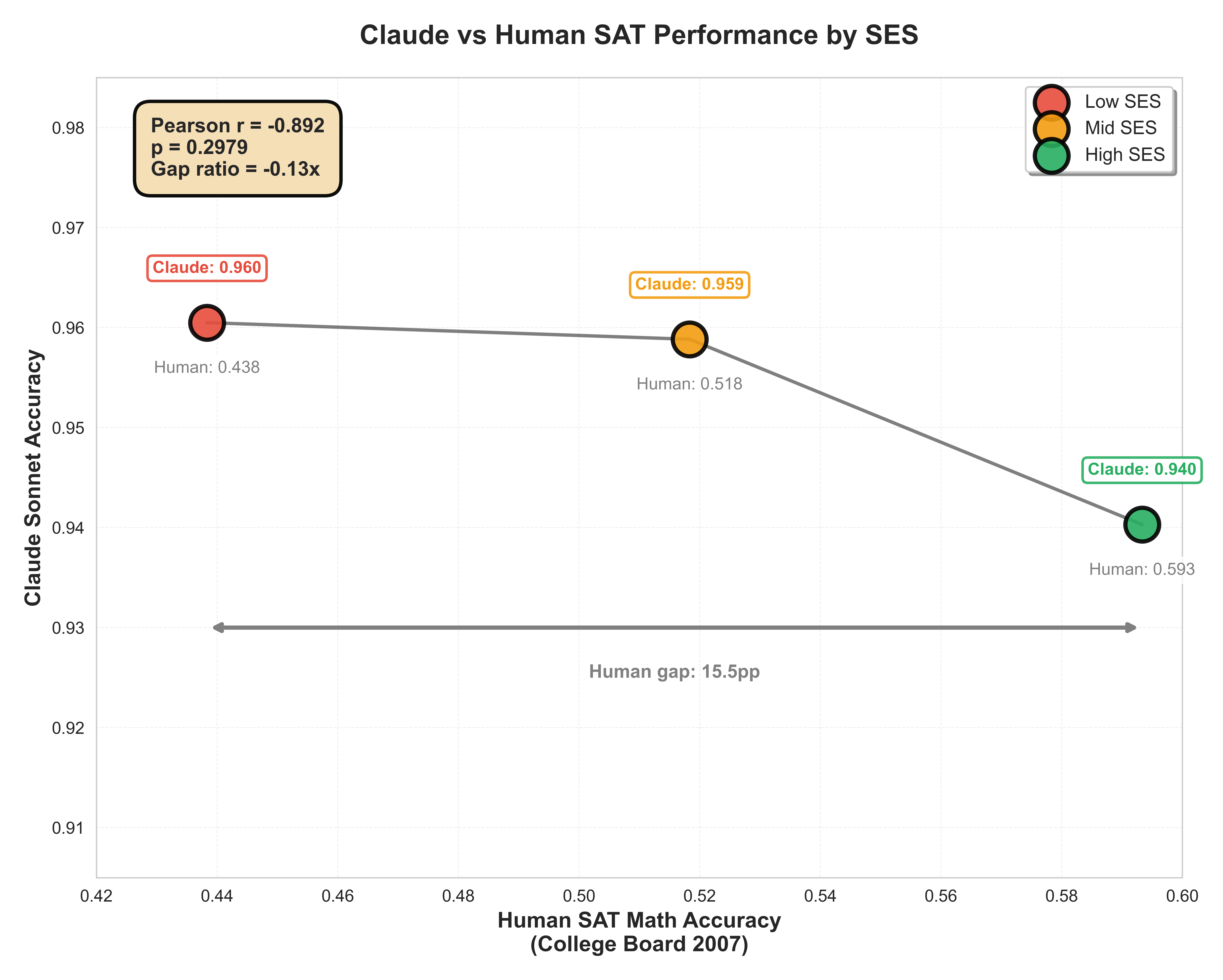}

    \caption{
      \textbf{AI vs.\ human SAT performance by socioeconomic background.}
      The plot compares Claude Sonnet agents' accuracy (Low vs.\ High SES personas)
      to human SAT math accuracy derived from College Board data.
      A paradoxical pattern emerges: \textbf{Low-SES Claude agents substantially outperform
        real Low-SES students}. This indicates that while Claude expresses SES-conditioned reasoning differences,
      it does not reproduce human-like performance gaps. Instead, the model is bounded by
      its training distribution, preventing realistic human-level trait embodiment or
      performance degradation among constrained personas.}
    \label{fig:ai_human_perf}
  \end{figure}

  \begin{figure}
    \includegraphics[width=0.9\textwidth]{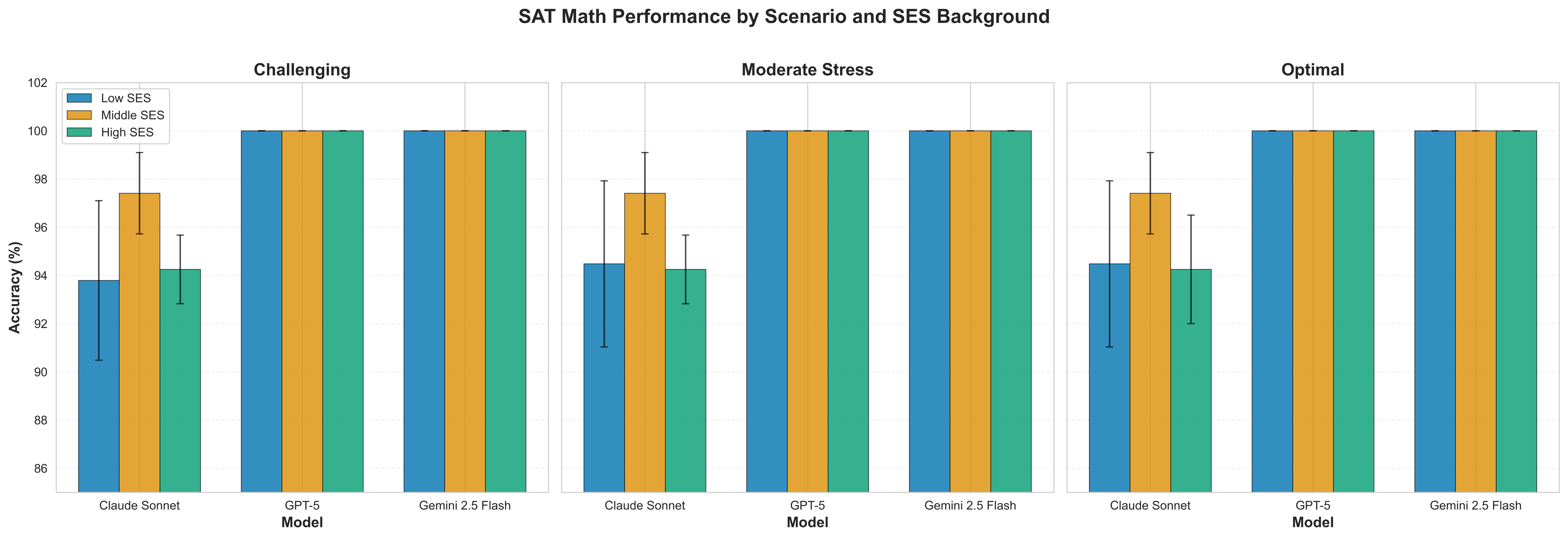}
    \captionof{figure}{SAT mathematics accuracy across socioeconomic personas and testing scenarios for each model for temperature \(T=0.6\):
      (a) GPT-5 exhibited complete contextual collapse with uniform accuracy across SES groups and scenarios.
      (b) Gemini 2.5 Flash also collapses under all scenarios and SES groups.
      (c) Claude Sonnet 4.5 retained measurable SES-based accuracy differences across all scenarios.}
    \label{fig:sat_accuracy_ses_scenarios06}
  \end{figure}

  \begin{figure}[ht]
    \centering
    \begin{subfigure}[t]{0.48\textwidth}
      \centering
      \includegraphics[width=\textwidth]{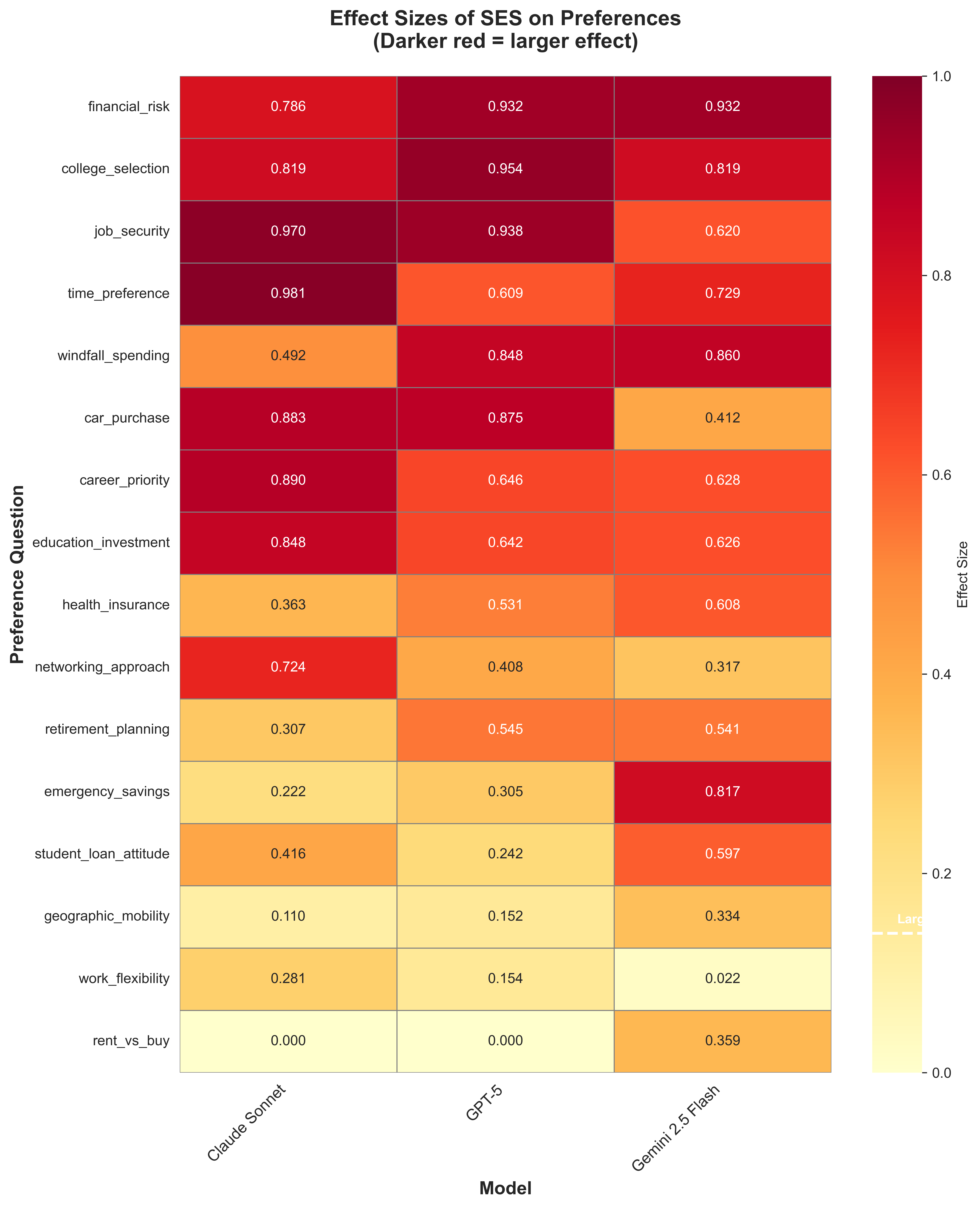}
      \caption{Effect size heatmap ($\epsilon^2$ / Cramer's $V$)}
      \label{fig:pref_effectsize_heatmap06}
    \end{subfigure}
    \hfill
    \begin{subfigure}[t]{0.48\textwidth}
      \centering
      \includegraphics[width=\textwidth]{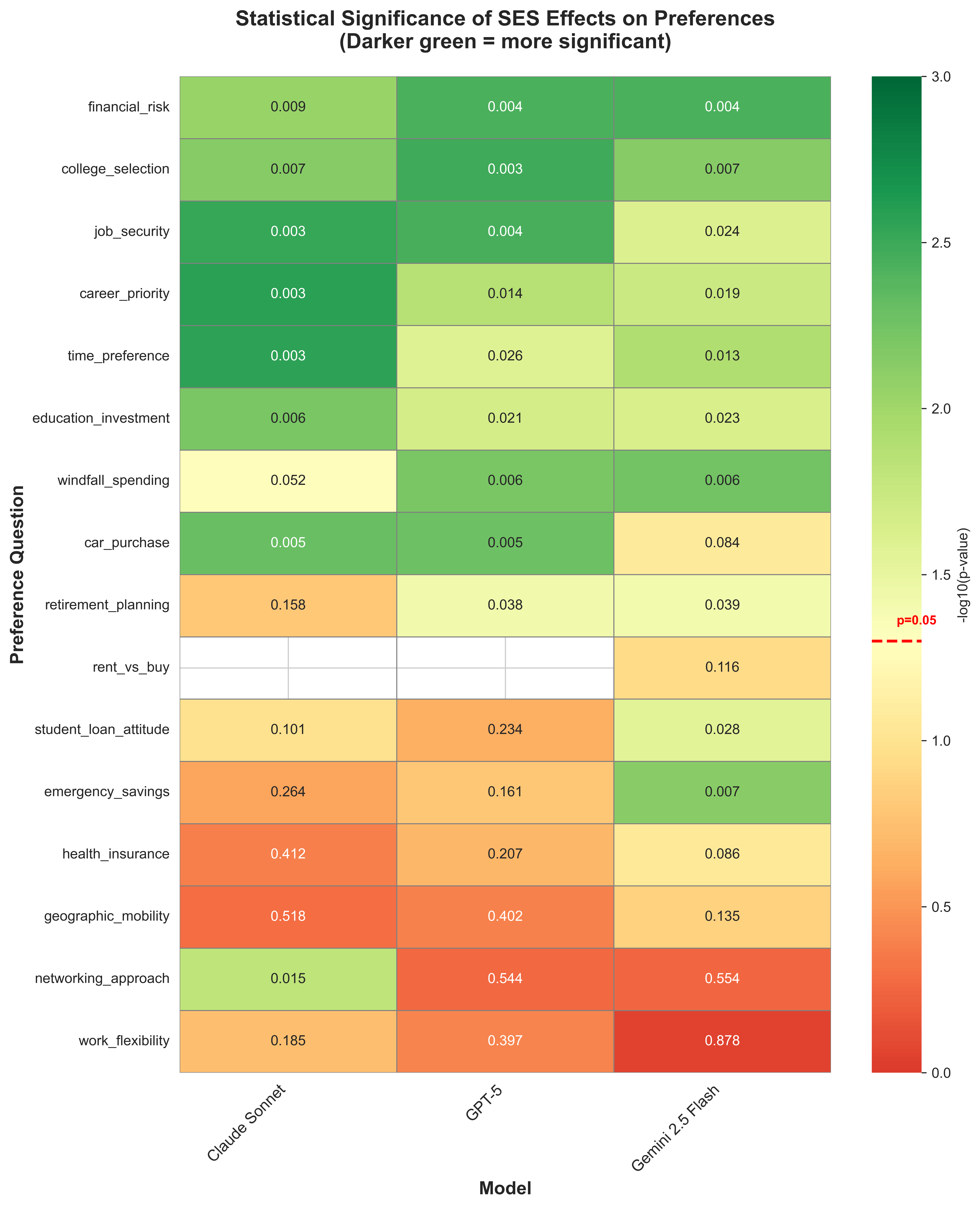}
      \caption{Statistical significance heatmap ($p$-values)}
      \label{fig:pref_pvalue_heatmap06}
    \end{subfigure}

    \caption{Preference task SES analysis across 16 economic items and three models for temperature \(T=0.6\):
      (a) Effect sizes for ordinal and categorical preference dimensions.
      (b) Corresponding $p$-value heatmap showing the robustness and direction of SES associations.}
    \label{fig:preference_heatmaps06}
  \end{figure}

  \begin{figure}[ht]
    \centering

    \begin{subfigure}[t]{0.48\linewidth}
      \centering
      \includegraphics[width=\linewidth]{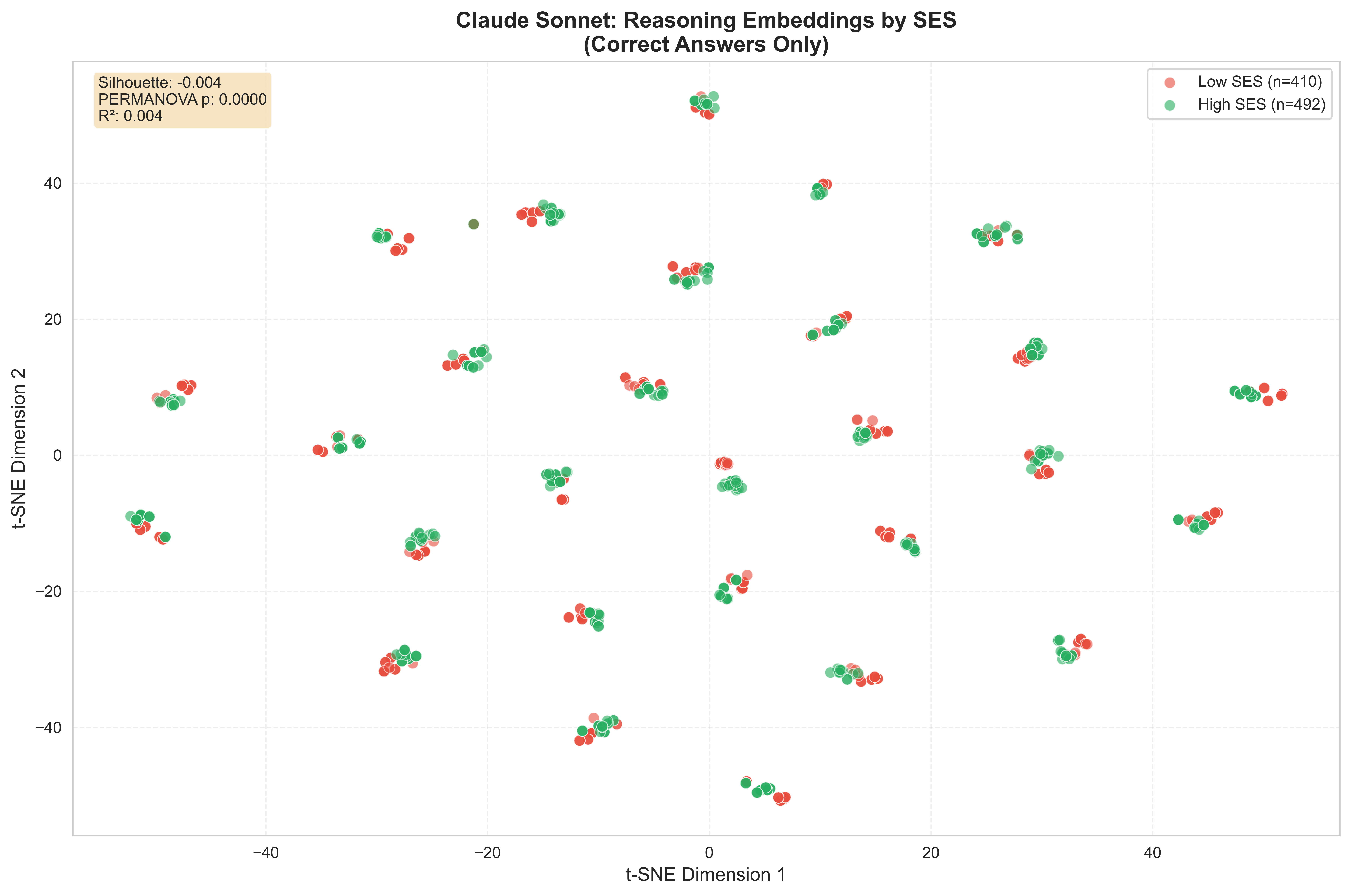}
      \caption*{\textbf{Claude Sonnet}\\\small Minor SES separation}
    \end{subfigure}
    \hspace{0.01\linewidth}
    \begin{subfigure}[t]{0.48\linewidth}
      \centering
      \includegraphics[width=\linewidth]{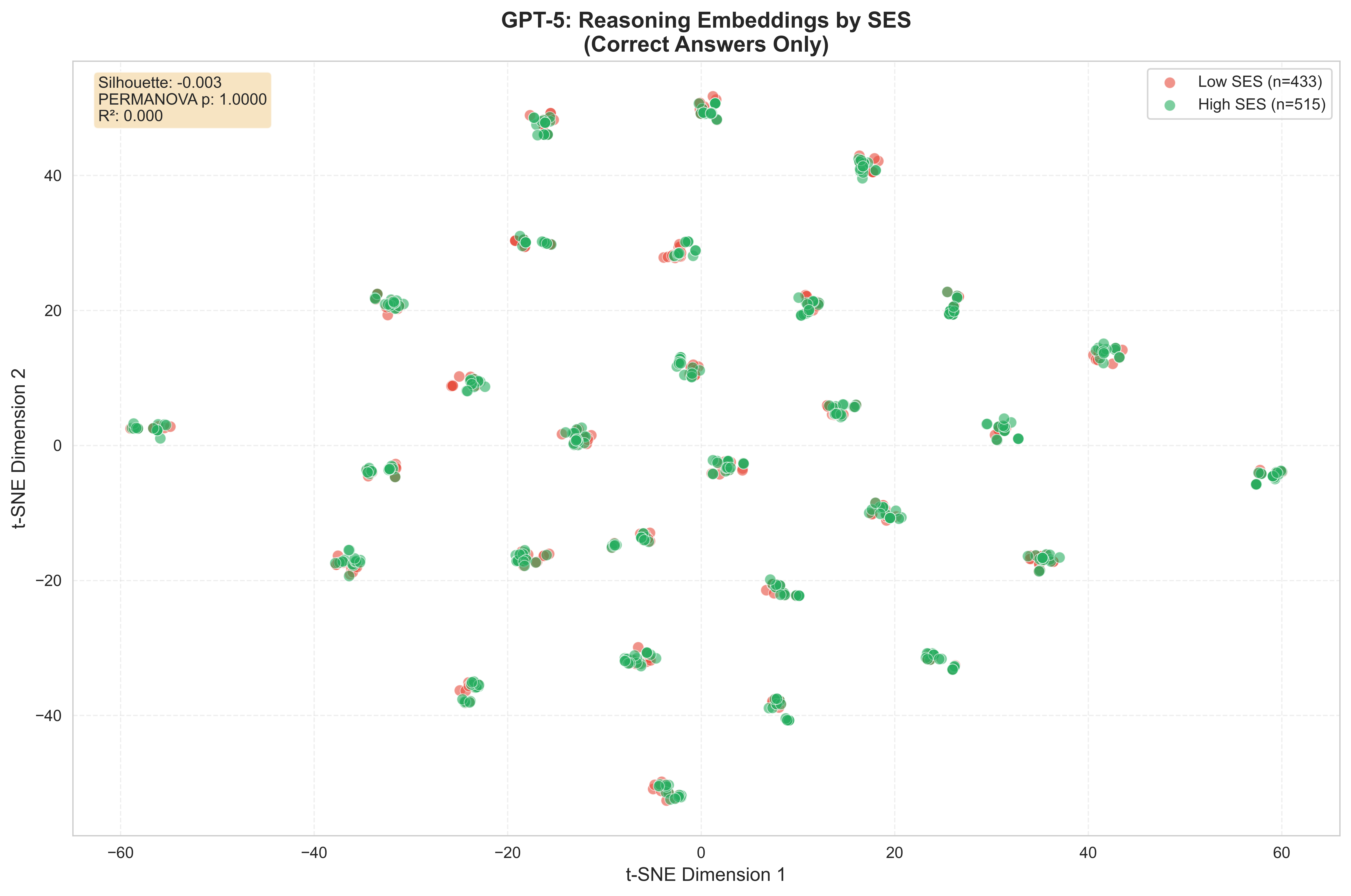}
      \caption*{\textbf{GPT-5}\\\small No SES separation}
    \end{subfigure}
    \hspace{0.01\linewidth}
    \begin{subfigure}[t]{0.48\linewidth}
      \centering
      \includegraphics[width=\linewidth]{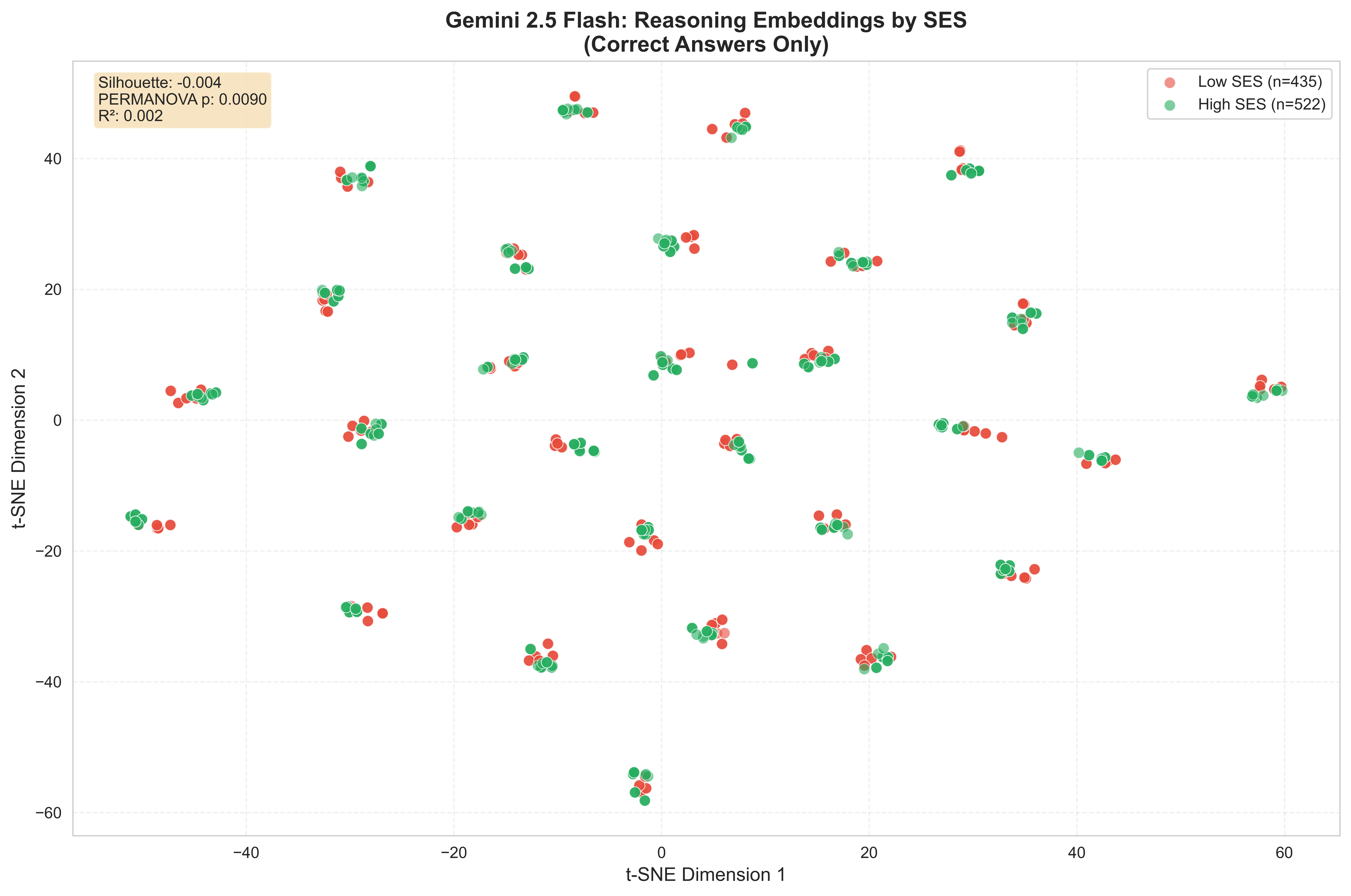}
      \caption*{\textbf{Gemini 2.5 Flash}\\\small Minor SES separation}
    \end{subfigure}

    \caption{
      t-SNE projections of reasoning embeddings from correct SAT solutions.
      Claude Sonnet and Gemini 2.5 exhibit slight SES-structured drift, while
      GPT-5 shows complete suppression of SES structure even for $T=0.6$.
    }
    \label{fig:ses_clustering_three_models06}
  \end{figure}

  \begin{figure}[ht]
    \centering

    \begin{subfigure}[t]{0.48\textwidth}
      \centering
      \includegraphics[width=\textwidth]{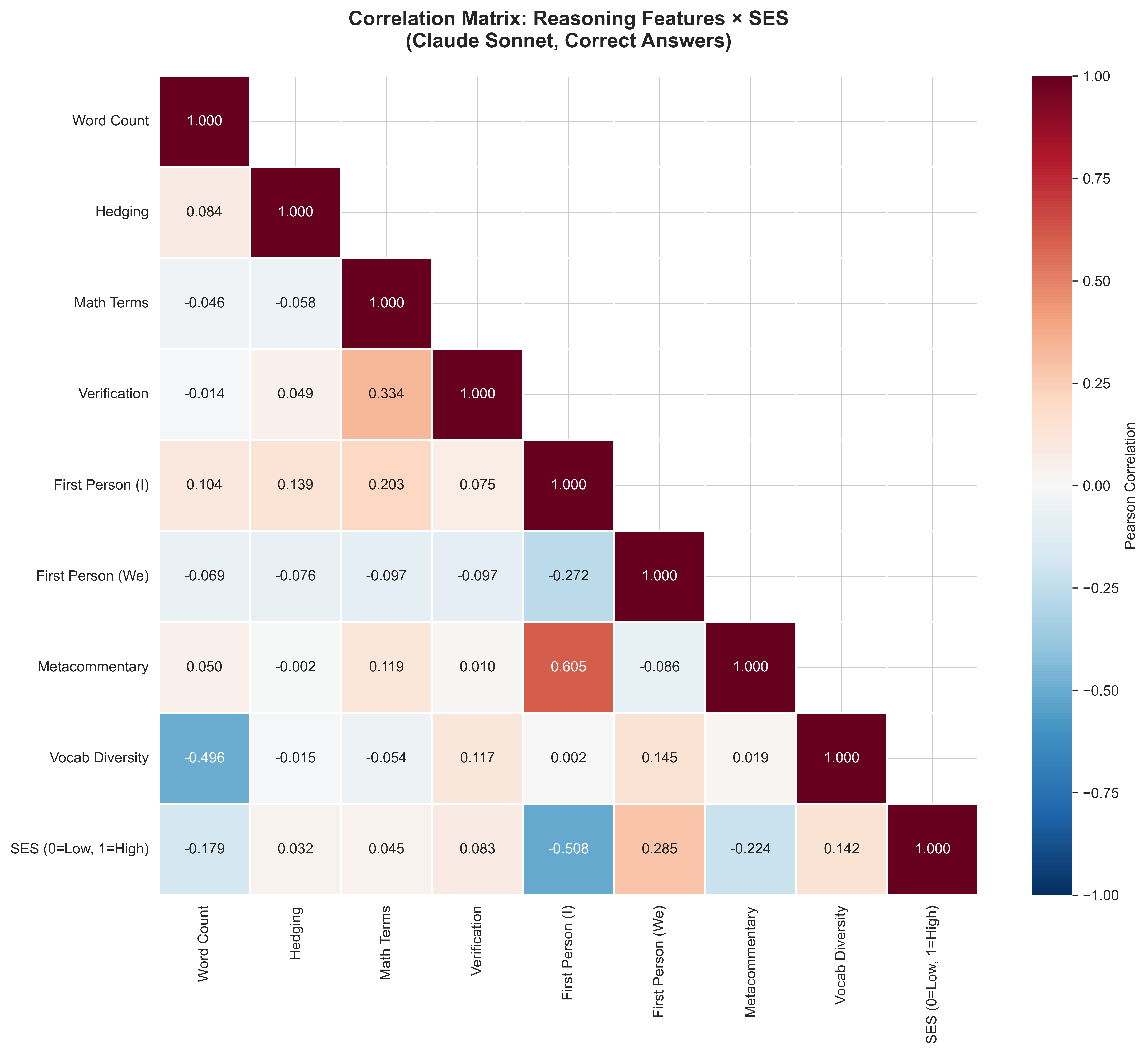}
      \caption{Linguistic feature correlation structure}
    \end{subfigure}
    \hfill
    \begin{subfigure}[t]{0.48\textwidth}
      \centering
      \includegraphics[width=\textwidth]{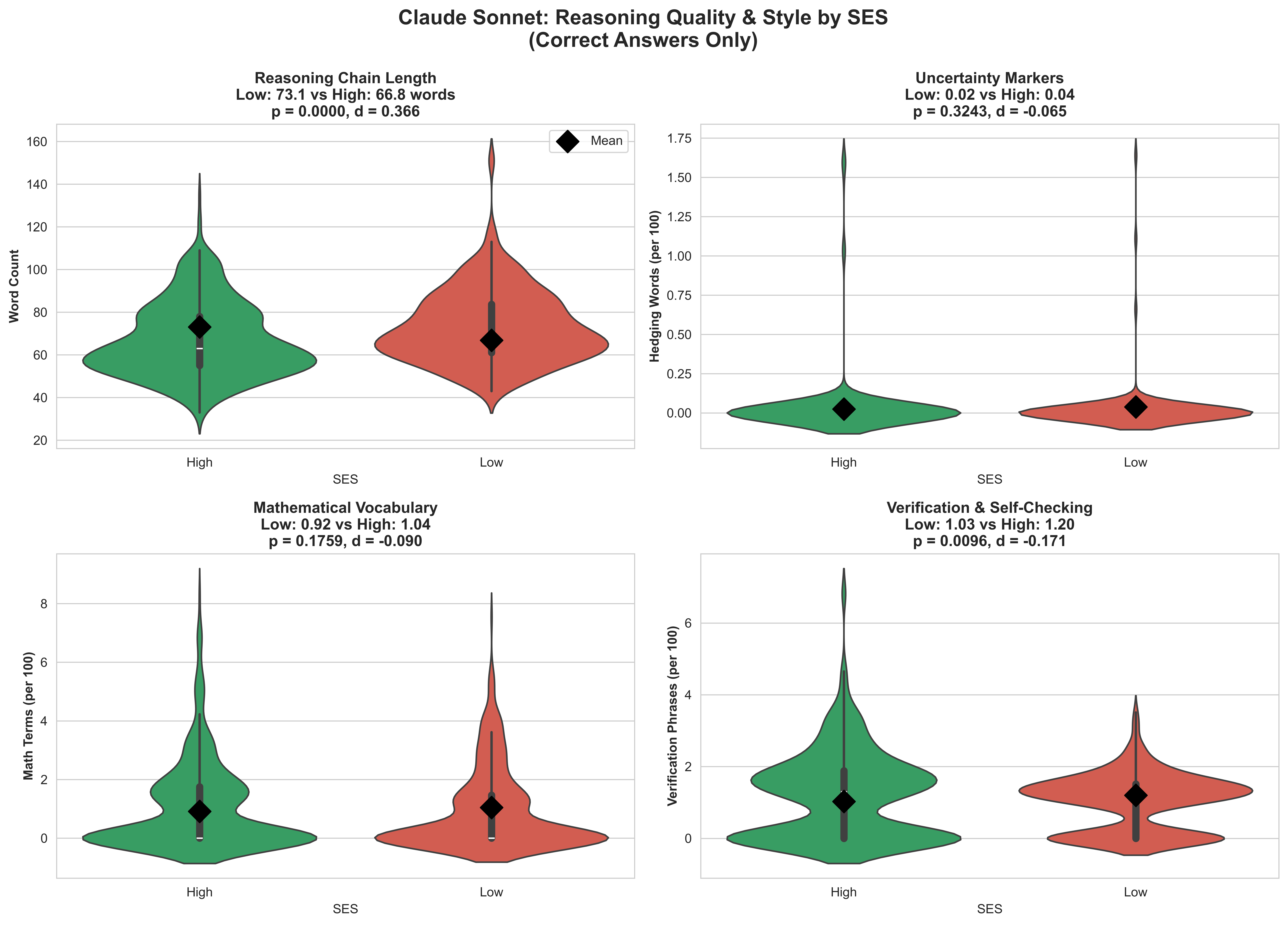}
      \caption{Top SES-differentiating linguistic features}
    \end{subfigure}

    \caption{Reasoning quality and linguistic structure in Claude compared to other models for temperature $T=0.6$.}
    \label{fig:reasoning_linguistic_results06}
  \end{figure}

  \begin{figure}[ht]
    \centering
    \includegraphics[width=0.85\textwidth]{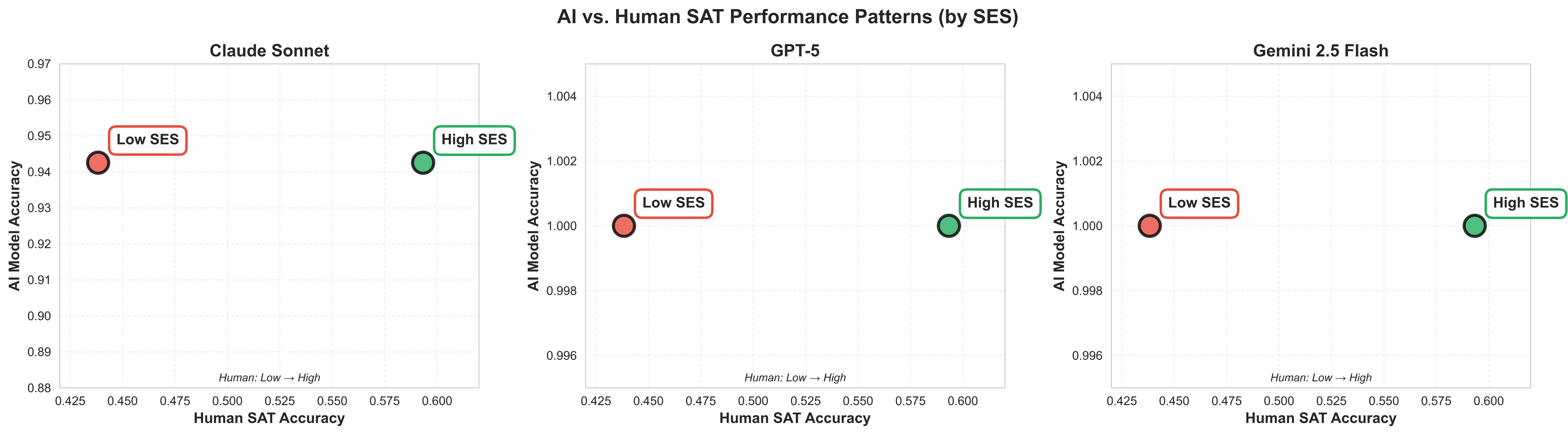}
    \caption{
      \textbf{Alignment between human SAT SES patterns and AI model SES patterns.}
      Scatter points show Low-SES and High-SES performance for humans and each model. The gradient for Claude in this context with temperature $T=0.6$ is zero, suggesting the main experiment results could be due to random variation. We additionally verify this in the Claude replication using extended agent list and recapture the inverted gradient with increased statistical power. Whereas GPT-5 and Gemini 2.5 Flash exhibit complete suppression of SES differences. The human accuracy is estimated using College Board reported data \cite{CollegeBoard2007} (see Section \ref{sec4.3.4}).}
    \label{fig:human_ai_alignment06}
  \end{figure}

  \begin{table}[ht]
    \centering
    \caption{\textbf{Low-SES personas (N = 15).}
      Personas reflect instability in schooling, limited resources, inconsistent test preparation, and elevated test anxiety.}
    \label{tab:personas_low}
    \vspace{0.5em}
    \renewcommand{\arraystretch}{1.25}
    \small
    \begin{tabular}{p{4.2cm}p{9.5cm}}
      \toprule
      \textbf{Persona Name}                      & \textbf{Summary Traits}                                   \\
      \midrule

      \textbf{migrant\_worker\_child (Carlos)}   &
      10th grader; highly mobile; major algebra gaps; unstable schooling; no prep; high anxiety.             \\[3pt]

      \textbf{reservation\_student (Aiyana)}     &
      11th grader from tribal school; limited STEM resources; self-study; cultural concerns about test bias. \\[3pt]

      \textbf{foster\_care\_student (Jordan)}    &
      12th grader with disrupted schooling; multiple school changes; limited online prep; high anxiety.      \\[3pt]

      \textbf{homeless\_student (Taylor)}        &
      11th grader experiencing homelessness; inconsistent attendance; free community prep when possible.     \\[3pt]

      \textbf{appalachian\_rural (Riley)}        &
      10th grader from rural Appalachia; limited advanced coursework; low-income; no local prep.             \\[3pt]

      \textbf{single\_parent\_household (Alex)}  &
      12th grader caring for siblings; limited study time; borrowed prep book; moderate–high anxiety.        \\[3pt]

      \textbf{recent\_refugee (Amina)}           &
      9th grader; strong math foundation but English barriers; nonprofit tutoring; very high anxiety.        \\[3pt]

      \textbf{rural\_farming\_community (Sam)}   &
      11th grader from farming town; solid basics; limited AP access; heavy work obligations.                \\[3pt]

      \textbf{teen\_parent (Morgan)}             &
      12th grader balancing childcare and school; declining performance; no prep; high anxiety.              \\[3pt]

      \textbf{unaccompanied\_minor (Diego)}      &
      10th grader with interrupted schooling; pro bono tutoring; legal stress; very high anxiety.            \\[3pt]

      \textbf{tribal\_urban\_relocatee (Kai)}    &
      11th grader; cultural displacement; average math; youth center tutoring; moderate anxiety.             \\[3pt]

      \textbf{chronic\_illness\_student (Jamie)} &
      12th grader with chronic illness; frequent absences; hospital tutoring; high stress.                   \\[3pt]

      \textbf{struggling\_rural\_student (Emma)} &
      10th grader from rural area; struggles with algebra; low-income; unfamiliar with SAT.                  \\[3pt]

      \textbf{urban\_underresourced (Marcus)}    &
      11th grader in underfunded school; limited prep; after-school work; moderate anxiety.                  \\[3pt]

      \textbf{ell\_student (Sofia)}              &
      9th-grade English-language learner; strong conceptual math; language-related test anxiety.             \\

      \bottomrule
    \end{tabular}
  \end{table}

  \begin{table}[ht]
    \centering
    \caption{\textbf{Middle-SES personas (N = 15).}
      Personas reflect educational stability, moderate resources, and varying levels of academic rigor and test preparation.}
    \label{tab:personas_middle}
    \vspace{0.5em}
    \renewcommand{\arraystretch}{1.25}
    \small
    \begin{tabular}{p{4.2cm}p{9.5cm}}
      \toprule
      \textbf{Persona Name}                         & \textbf{Summary Traits}                               \\
      \midrule

      \textbf{military\_dependent (Casey)}          &
      11th grader who moves frequently; solid math; DoDEA exposure; SAT workshops; moderate anxiety.        \\[3pt]

      \textbf{working\_class\_striving (Pat)}       &
      10th grader from trade-skill family; hardworking B student; paid community SAT course; part-time job. \\[3pt]

      \textbf{suburban\_commuter (Drew)}            &
      11th grader from commuter suburb; average math; group prep class; low–moderate anxiety.               \\[3pt]

      \textbf{arts\_focused\_student (Skyler)}      &
      12th grader strong in arts; weaker in math; short targeted prep; math-specific anxiety.               \\[3pt]

      \textbf{dual\_enrollment\_student (Avery)}    &
      11th grader taking college courses; above-average math; self-studied; moderate anxiety.               \\[3pt]

      \textbf{homeschool\_student (River)}          &
      10th grader; independent learning; some gaps; online prep; moderate anxiety.                          \\[3pt]

      \textbf{first\_gen\_immigrant (Lin)}          &
      11th grader; strong math; free community prep; intense family expectations.                           \\[3pt]

      \textbf{religious\_school\_student (Gabriel)} &
      12th grader at modest private school; consistent B+; in-school SAT prep.                              \\[3pt]

      \textbf{small\_town\_athlete (Quinn)}         &
      11th grader motivated by NCAA eligibility; average math; online prep; moderate anxiety.               \\[3pt]

      \textbf{career\_tech\_pathway (Reese)}        &
      12th grader in CTE track; strong applied math; basic school prep; moderate anxiety.                   \\[3pt]

      \textbf{distracted\_suburban (Tyler)}         &
      10th grader; capable but inattentive; minimal prep; low motivation.                                   \\[3pt]

      \textbf{rural\_average (Hannah)}              &
      11th grader from small rural town; working-class; self-studies via Khan Academy.                      \\[3pt]

      \textbf{typical\_suburban (Ethan)}            &
      11th grader; B-level math; summer SAT prep; moderate anxiety.                                         \\[3pt]

      \textbf{urban\_motivated (Aisha)}             &
      11th grader; strong commitment; heavy use of free prep; high expectations.                            \\[3pt]

      \textbf{small\_town\_steady (Noah)}           &
      12th grader; consistent B+; limited AP access; low–moderate anxiety.                                  \\

      \bottomrule
    \end{tabular}
  \end{table}

  \begin{table}[ht]
    \centering
    \caption{\textbf{High-SES personas (N = 15).}
      Personas reflect enriched educational environments, extensive tutoring resources, and strong college-oriented preparation.}
    \label{tab:personas_high}
    \vspace{0.5em}
    \renewcommand{\arraystretch}{1.25}
    \small
    \begin{tabular}{p{4.2cm}p{9.5cm}}
      \toprule
      \textbf{Persona Name}                           & \textbf{Summary Traits}                    \\
      \midrule

      \textbf{legacy\_ivy\_aspirant (Harper)}         &
      12th grader in elite school; exceptional math; extensive private tutoring; very low anxiety. \\[3pt]

      \textbf{international\_school\_student (Priya)} &
      11th grader in rigorous international school; IB curriculum; professional test prep.         \\[3pt]

      \textbf{silicon\_valley\_prodigy (Kai)}         &
      10th grader in tech hub; advanced STEM; coding competitions; minimal prep needed.            \\[3pt]

      \textbf{prep\_school\_athlete (Blake)}          &
      12th grader at top prep school; strong all-around performance; integrated tutoring.          \\[3pt]

      \textbf{medical\_family\_achiever (Sophia)}     &
      11th grader; excellent STEM ability; physician parents; private tutoring; high expectations. \\[3pt]

      \textbf{arts\_and\_academics\_elite (Julian)}   &
      12th grader strong in both arts and STEM; conservatory-level training; balanced prep.        \\[3pt]

      \textbf{debate\_champion (Cameron)}             &
      11th grader; nationally-ranked debater; strong reasoning; targeted test prep.                \\[3pt]

      \textbf{summer\_program\_circuit (Madison)}     &
      11th grader attending elite programs (RSI/TASP); extensive enrichment; private tutoring.     \\[3pt]

      \textbf{college\_prep\_achiever (Olivia)}       &
      11th grader; AP Calculus; multiple enrichment activities; private SAT tutor.                 \\[3pt]

      \textbf{magnet\_school\_star (Jayden)}          &
      12th grader in selective STEM magnet; advanced curriculum; robotics captain.                 \\[3pt]

      \textbf{motivated\_upward\_mobility (Maya)}     &
      12th grader; self-driven; extensive free prep; pursuing scholarships.                        \\[3pt]

      \textbf{private\_school\_elite (Alexander)}     &
      12th grader in elite private school; affluent family; 50+ hours tutoring.                    \\[3pt]

      \textbf{math\_competition\_specialist (Daniel)} &
      11th grader; AMC/AIME competitor; minimal formal prep; high intrinsic motivation.            \\[3pt]

      \textbf{prodigy\_accelerated (Zoe)}             &
      10th grader taking university-level math; early-identified gifted student.                   \\[3pt]

      \textbf{well\_rounded\_achiever (Liam)}         &
      12th grader excelling across subjects; comprehensive prep; upper-middle-class family.        \\

      \bottomrule
    \end{tabular}
  \end{table}

  \begin{table}[ht]
    \centering
    \caption{
      \textbf{SES Differences in SAT Accuracy Across Scenarios (15-Agent Replication) using $T=0.6$.}
      Claude Sonnet shows small SES-linked variation across scenarios, whereas
      GPT-5 and Gemini 2.5 Flash show zero variance (ceiling performance).
    }
    \label{tab:anova_replication}
    \begin{tabular}{l l c c c c c c}
      \toprule
      \textbf{Model}   & \textbf{Scenario} & \textbf{F}   & \textbf{p}    & $\boldsymbol{\eta^2}$
                       & \textbf{Low}      & \textbf{Mid} & \textbf{High}                                                 \\
      \midrule
      Claude Sonnet    & Challenging       & 2.48         & 0.125         & 0.293                 & 0.938 & 0.974 & 0.943 \\
      Claude Sonnet    & Moderate Stress   & 1.91         & 0.191         & 0.241                 & 0.945 & 0.974 & 0.943 \\
      Claude Sonnet    & Optimal           & 1.50         & 0.263         & 0.200                 & 0.945 & 0.974 & 0.943 \\
      \midrule
      GPT-5            & Challenging       & --           & --            & 0                     & 1.000 & 1.000 & 1.000 \\
      GPT-5            & Moderate Stress   & --           & --            & 0                     & 1.000 & 1.000 & 1.000 \\
      GPT-5            & Optimal           & --           & --            & 0                     & 1.000 & 1.000 & 1.000 \\
      \midrule
      Gemini 2.5 Flash & Challenging       & --           & --            & 0                     & 1.000 & 1.000 & 1.000 \\
      Gemini 2.5 Flash & Moderate Stress   & --           & --            & 0                     & 1.000 & 1.000 & 1.000 \\
      Gemini 2.5 Flash & Optimal           & --           & --            & 0                     & 1.000 & 1.000 & 1.000 \\
      \bottomrule
    \end{tabular}
  \end{table}

\end{appendix}

\end{document}